\documentclass[twocolumn]{aastex631}
\usepackage{amsmath}
\usepackage{natbib}
\usepackage{graphicx}
\usepackage{url}

\usepackage{verbatim}

\shorttitle{Simulating Disk Ices}
\shortauthors{Ballering, Cleeves, \& Anderson}

\begin{document}

\title{Simulating Observations of Ices in Protoplanetary Disks}

\author[0000-0002-4276-3730]{Nicholas P. Ballering}
\altaffiliation{Virginia Initiative on Cosmic Origins Fellow.}
\affiliation{Department of Astronomy, University of Virginia, Charlottesville, VA 22904, USA}

\author[0000-0003-2076-8001]{L. Ilsedore Cleeves}
\affiliation{Department of Astronomy, University of Virginia, Charlottesville, VA 22904, USA}
\affiliation{Department of Chemistry, University of Virginia, Charlottesville, VA 22904, USA}

\author[0000-0002-8310-0554]{Dana E. Anderson}
\altaffiliation{Virginia Initiative on Cosmic Origins Fellow.}
\affiliation{Department of Astronomy, University of Virginia, Charlottesville, VA 22904, USA}

\correspondingauthor{Nicholas P. Ballering}
\email{nb2ke@virginia.edu}

\received{2021 May 25}
\revised{2021 July 14}
\accepted{2021 July 19}

\begin{abstract}
Ices are an important constituent of protoplanetary disks. New observational facilities, notably the James Webb Space Telescope (JWST), will greatly enhance our view of disk ices by measuring their infrared spectral features. We present a suite of models to complement these upcoming observations. Our models use a kinetics-based gas-grain chemical evolution code to simulate the distribution of ices in a disk, followed by a radiative transfer code using a subset of key ice species to simulate the observations. We present models reflecting both molecular inheritance and chemical reset initial conditions. We find that near-to-mid-IR absorption features of H$_2$O, CO$_2$, and CH$_3$OH are readily observable in disk-integrated spectra of highly inclined disks while CO, NH$_3$, and CH$_4$ ice do not show prominent features. CH$_3$OH ice has low abundance and is not observable in the reset model, making this species an excellent diagnostic of initial chemical conditions. CO$_2$ ice features exhibit the greatest change over disk lifetime, decreasing and increasing for the inheritance and reset models, respectively. Spatially resolved spectra of edge-on disks, possible with JWST's integral field unit observing modes, are ideal for constraining the vertical distribution of ices and may be able to isolate features from ices closer to the midplane (e.g., CO) given sufficient sensitivity. Spatially-resolved spectra of face-on disks can trace scattered-light features from H$_2$O, CO$_2$, and CH$_3$OH, plus CO and CH$_4$ from the outermost regions. We additionally simulate far-IR H$_2$O ice emission features and find they are strongest for disks viewed face-on.                            
\end{abstract}

\section{INTRODUCTION}
\label{sec:introduction}

Protoplanetary disks consist of molecular and atomic gases, refractory dust, and volatile ices. Ices play a critical role in the planet formation process. They increase the sticking efficiency of dust grains \citep{wang2005,gundlach2015}, facilitating their growth to larger pebbles (but see also, e.g.,  \citealt{musiolik2016}). A snow line---the location in the disk midplane where a given species transitions from gas to ice---can promote the growth of planetesimals by enhancing the local concentration of disk material \citep[e.g.,][]{stevenson1988,drazkowska2017}. 

Ices are crucial for the delivery of biocritical volatiles to potentially habitable planets. Terrestrial planets in the habitable zone are thought to form within the snow lines of water and other volatile species, leaving them dry and unsuitable for life. Volatiles must be delivered to these planets from ice-containing planetesimals that formed farther out in the disk. This could occur from planetesimal scattering during the final assembly of giant planets \citep[e.g.,][]{raymond2017} or from later epochs of dynamical instability \citep[e.g.,][]{gomes2005}.

Studies of dust \citep{andrews2020} and gas \citep[e.g.,][]{pontoppidan2014,facchini2021,oberg2021} have revealed much about the overall physical and chemical properties of disks, yet fundamental questions remain unanswered \citep[e.g.,][]{haworth2016,oberg2021_review}. Studying ices offers a unique and complementary opportunity to probe disk properties, such as their temperature structure and internal radiation field, the sizes of dust grains in various disk regions, the degree to which the molecular inventory is inherited from the protostellar envelope or altered during disk formation, and the amount of vertical and radial mixing of disk material. Comprehensive disk models have been developed to simultaneously fit panchromatic observations of dust and gas \citep[e.g.,][]{woitke2019}; the inclusion of ice observations can further constrain such models.    

Interstellar ices are detected via spectral features in absorption and scattering at near-to-mid-IR wavelengths \citep{boogert2015}. H$_2$O ice also exhibits emission features at 44 and 62 $\micron$. Ices have been studied across all star- and planet-forming environments, including molecular clouds \citep[e.g.,][]{whittet1983,bergin2005}, dense cores \citep[e.g.,][]{boogert2011}, protostellar envelopes \citep[e.g.,][]{boogert2008,oberg2011}, and in comets/icy planetesimals both in our solar system \citep{mumma2011,altwegg2019} and around other stars \citep{matra2017,matra2018,strom2020}.

Protoplanetary disks are the evolutionary link between molecular clouds and planetesimals and planets, yet there are few observational constraints on the abundance and distribution of ices in disks. Absorption spectroscopy is only possible when a disk is viewed close to edge-on such that the star and hot inner disk provide a background source of radiation observed through the cooler outer disk where the ices reside. The dense disk midplane occults the star and warm inner disk, so edge-on systems are faint in the near-to-mid-IR and most of the observed flux is due to scattered light from dust above the midplane. This faintness has limited the detectability of ice to especially bright disks. H$_2$O ice was detected from several disks via its 3 $\micron$ absorption feature using ground-based telescopes \citep[e.g.][]{terada2007,mccabe2011,terada2012,terada2017}. Ices besides H$_2$O have only been detected from a few disks with AKARI \citep{aikawa2012} and Spitzer \citep{pontoppidan2005}, although the analysis was complicated by potential foreground contamination of the features, external to the disk itself. 

Ices can also be detected via scattered-light features \citep{inoue2008}. To date, the 3 $\micron$ H$_2$O feature has been successfully detected in a few disks using photometric imaging. Images are taken at wavelengths spanning the 3 $\micron$ feature and show darkening at 3 $\micron$ compared to adjacent wavelengths, indicative of the reduced albedo due to the presence of water ice \citep{honda2009,honda2016,betti2021}.

Finally, the H$_2$O far-IR emission features were detected from a handful of disks with the Infrared Space Observatory \citep{malfait1998,malfait1999,vandenancker2000,creech-eakman2002} and the Herschel Space Observatory \citep{mcclure2015,min2016}. The shapes of the far-IR features are particularly sensitive to crystallinity \citep{kamp2018}. 

Upcoming observational facilities are poised to revolutionize our view of ices in disks. The James Webb Space Telescope (JWST), with a wavelength range from 0.6 to 28.8 $\mu m$, will cover all known absorption/scattering ice features. The JWST NIRSpec and MIRI spectrographs will offer unprecedented sensitivity and higher spectral resolution than previous IR facilities. These spectrographs also have integral field unit (IFU) capabilities, allowing the location of the ice features to be mapped for spatially resolved disks. The NIRCAM and MIRI instruments offer coronagraphy, which can photometrically  image ice via scattered light.

Beyond JWST, the Spectro-Photometer for the History of the Universe, Epoch of Reionization, and Ices Explorer (SPHEREx), scheduled for launch in 2024, will acquire 0.75--5.0~$\mu m$ spectrum over the whole sky \citep{dore2018}. This wavelength range covers features of H$_2$O, CO, CO$_2$, CH$_3$OH, NH$_3$, and CH$_4$ ices, among others. While SPHEREx will not have the same sensitivity, spectral resolution, or spatial resolution as JWST, it will observe many more disks. New far-IR observing capabilities will be important for detecting the H$_2$O emission features. Such capabilities may arrive from new instrumentation aboard the Stratospheric Observatory for Infrared Astronomy (SOFIA) or from the proposed Origins Space Telescope.

To complement upcoming observations, we present simulated observations of the spectral features of disk ices based on detailed chemical evolution models. Previous studies have used radiative transfer simulations to interpret/predict ice features, but they did not make use of disk chemical evolution models to set up the ice abundance distributions. \citet{pontoppidan2005} interpreted edge-on disk observations with a model that included CO ice where the temperature was below 20 K and CO$_2$ and H$_2$O ice where the temperature was below 90 K. \citet{mcclure2015} and \citet{min2016} interpreted far-IR H$_2$O features using a model with total ice mass as a free parameter and its distribution set by local temperature and UV extinction. \citet{kamp2018} also used this approach to explore the general behavior of the far-IR H$_2$O features under various disk conditions. A similar model was developed by \citet{oka2012} and employed by \citet{honda2016} to analyze the 3 $\micron$ scattered-light feature of H$_2$O ice. 

The clear next step is to incorporate our knowledge of evolving disk chemistry into our predictions for ice observations, which is the goal this paper sets out to achieve. The chemical and radiative transfer models are described in Section \ref{sec:methods}. Section \ref{sec:results} presents the results, highlighting the distribution and evolution of ice abundances (\ref{sec:abundances}), the effect of disk inclination (\ref{sec:inclination}), the spatial variation of the features (\ref{sec:spatial}), variations in the features with time and initial disk chemistry (\ref{sec:intialandtime}), and detectability with JWST (\ref{sec:JWST}). The conclusions of this study are summarized in Section \ref{sec:summary}.

\section{Methods}
\label{sec:methods}

\begin{figure*}
\includegraphics[width=1\textwidth]{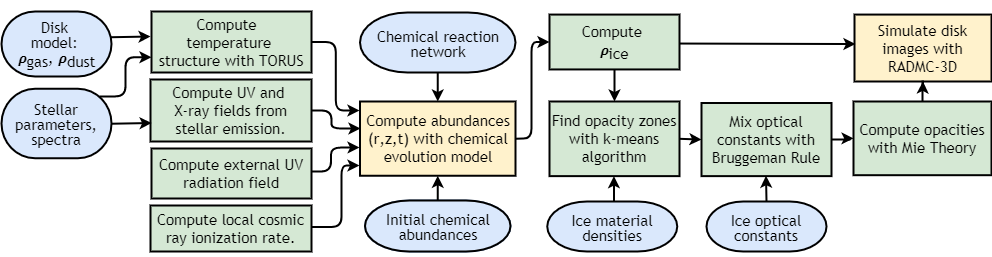}
\caption{Flowchart of the modeling procedure. Blue ovals indicate model inputs, green squares indicate various computational steps, and yellow squares indicate the steps that yield our primary model outputs: ice abundances and simulated observations.}
\label{fig:flowchart}
\end{figure*}

Our modeling procedure involves two main stages. First, we use a chemical evolution code to simulate the abundance of many species in gas and ice phases as a function of radial location, vertical location, and time. Second, we use radiative transfer to simulate observations of the dust plus a subset of ice species in the disk at four selected times. The specific steps of our modeling procedure are outlined in Figure \ref{fig:flowchart} and detailed in the following sections.

\subsection{Disk Properties}

\begin{deluxetable}{lcc}
\tablewidth{0pt}
\tablecolumns{3}
\tablecaption{Model Parameters \label{table:parameters}}
\tablehead{\colhead{Symbol} & \colhead{Description} &  \colhead{Value}}
\startdata
$r_\text{in}$ & Inner radius & 0.1 au \\
$r_c$ & Characteristic size & 30 au \\
$r_\text{out}$ & Outer radius & 120 au \\
$\Sigma_c$ & Characteristic surface density & 23 g cm$^{-2}$ \\ 
$\gamma$ & Radial power law & 1.0 \\
$M_d$ & Disk mass & 0.014 $M_\sun$ \\ 
$H_\text{100}$ & Scale height at 100 au & 10 au \\
$\psi$ & Flaring parameter & 1.15 \\
$\xi$ & Total dust-to-gas mass ratio & 0.01 \\
$f$ & Large dust mass fraction  & 0.9 \\
$\chi$ & Large dust settling factor & 0.2 \\
$a_\text{min}$ & Minimum grain size & 0.005 $\micron$ \\
$a_\text{max,s}$ & Small dust maximum grain size & 1 $\micron$ \\
$a_\text{max,l}$ & Large dust maximum grain size & 1000 $\micron$ \\
$q$ & Grain size distribution index & 3.5 \\
$T_\star$ & Stellar effective temperature & 4300 K \\
$R_\star$ & Stellar radius & 2.804 $R_\sun$ \\
$\zeta_\text{CR}$ & Incident CR ionization rate & $1.1 \times 10^{-18}$ s$^{-1}$ \\ 
\enddata
\end{deluxetable}

Our model was designed to represent a typical T Tauri class II disk system. We assume an azimuthally symmetric disk with a gas surface density profile given by
\begin{equation}
\label{eq:sigma}
\Sigma_g(r)=\Sigma_c\left(\frac{r}{r_c}\right)^{-\gamma}\exp\left[-\left(\frac{r}{r_c}\right)^{2-\gamma}\right]
\end{equation}
within $r_\text{in}<r<r_\text{out}$, as predicted by accretion theory \citep{lynden-bell1974,hartmann1998}. The characteristic disk size is $r_c$ = 30 au where the characteristic disk gas surface density is $\Sigma_c$ = 23 g cm$^{-2}$, which sets the total disk mass (0.014 $M_\sun$). The disk has a flared vertical structure with scale height
\begin{equation}
H(r) = H_\text{100}\left(\frac{r}{\text{100 au}}\right)^\psi,
\end{equation} where we set $H_\text{100}$ = 10 au and $\psi$ = 1.15. We assume an overall dust-to-gas mass ratio of $\xi=0.01$. The dust is partitioned into a small-grain (0.005--1 $\micron$) population and a large-grain (0.005--1000 $\micron$) population with a mass fraction of $f=0.9$ in the large grains. The small grains trace the same spatial distribution as the gas with volume density profile
\begin{equation}
\label{eq:smalldensity}
\rho_s(r,z) = \frac{\xi(1-f)\Sigma_g(r)}{\sqrt{2\pi} H(r)} \exp\left[-\frac{1}{2}\left(\frac{z}{H(r)}\right)^2\right].
\end{equation}
The large grains have the same radial distribution as the gas and small grains but a settled vertical distribution with the scale height modified by $\chi=0.2$. The large-grain volume density is thus
\begin{equation}
\label{eq:largedensity}
\rho_l(r,z) = \frac{\xi f\Sigma_g(r)}{\sqrt{2\pi}\chi H(r)} \exp\left[-\frac{1}{2}\left(\frac{z}{\chi H(r)}\right)^2\right].
\end{equation}
For both populations, the grain size distribution follows a power law, $n(a) \propto a^{-q}$ with $q=$3.5 \citep{mathis1977}, and the dust composition is 80\% silicates and 20\% graphite. This disk parameterization has been used by many previous studies when modeling disk structure and chemistry \citep[e.g.,][]{andrews2011,cleeves2013,cleeves2014_cr}. Table \ref{table:parameters} summarizes the model parameters.

\subsection{Disk Temperature and Radiation Field}
\label{sec:tempandradfield}

We compute the dust temperature and radiation field throughout the disk---all of which remain fixed during the subsequent simulation of chemical evolution---using the TORUS radiative transfer code \citep{harries2000,harries2004,kurosawa2004,pinte2009}. The central star has $T_\star$ = 4300 K and $R_\star$ = 2.804 $R_\sun$. We use an ATLAS9 model \citep{castelli2003} for the spectrum of the stellar photosphere.

The UV radiation from T Tauri stars is dominated by accretion luminosity \citep{calvet98,gullbring2000} while the X-ray radiation originates in the stellar corona and/or accretion shock \citep{feigelson1999,preibisch2005}, and for these we use a template spectrum based on observations of TW Hya  \citep{cleeves2013}. We use the Monte Carlo radiation transfer code of \citet{bethell2011_photoe,bethell2011_lya} to simulate the propagation of the UV and X-ray radiation in the disk, accounting for absorption and scattering by dust and resonant scattering of Ly$\alpha$ photons by atomic hydrogen. We allow the gas to be heated above the dust temperature by UV-driven processes. The interstellar UV field is not included in this model. Cosmic rays are another source of ionization, although stellar winds and magnetic fields may significantly attenuate the cosmic-ray (CR) flux \citep{cleeves2013,cleeves2014_cr}. We include an incident CR ionization rate of $\zeta_\text{CR}$ = 1.1$\times$10$^{-18}$ s$^{-1}$ per H$_2$ that is attenuated with vertical depth in the disk, motivated by observational constraints \citep{cleeves2015_TWHya,seifert2021}. This value of $\zeta_\text{CR}$ is at the lower end of the range typically explored in other chemical evolution models \citep{bosman2018,krijt2020}.

\subsection{Chemical Evolution Code}

The chemical model is initialized with a uniform set of initial abundances. We explore two sets of initial conditions to reflect two extreme assumptions: molecular cloud-like composition (Inheritance) and primarily atomic composition (Reset). In the case of Reset, an exception was made for hydrogen, which is molecular rather atomic in the initial conditions. The initial abundances for each chemical model are listed in Table \ref{table:initialabundances}.

For the Inheritance model, the H$_2$O ice abundance was set to 8 $\times$ 10$^{-5}$ relative to the total number of H atoms. This is near the high end of low-mass protostellar abundance measurements, as compiled by \citet{boogert2015}. We chose the high end because the measurements are likely biased to lower values due to optical depth effects caused by grain growth. The abundances of CO$_2$, NH$_3$, CH$_3$OH, and CH$_4$ ices were set based on the average of their measurements relative to  H$_2$O ice \citep{boogert2015}. The total nitrogen abundance measured in the diffuse interstellar medium (ISM) is 7.5 $\times$ 10$^{-5}$ \citep{meyer1997}, and we subtracted from this the nitrogen in NH$_3$ to derive the N$_2$ abundance. The CO abundance was derived from the ISM CO abundance of 1.3 $\times$ 10$^{-4}$ \citep{ripple2013} minus the measured carbon abundance in CO$_2$. The abundances of HCN, H$_3^+$, HCO$^+$, C$_2$H, SO, CS, Si$^+$, Mg$^+$, and Fe$^+$ are adopted from \citet{cleeves2014_cr}.

We evolve the abundances forward in time using the code of \cite{fogel2011} and \citet{cleeves2014_cr}, with updates to the grain surface chemistry, temperature-dependent sticking coefficients, and N$_2$ self-shielding by \citet{cleeves2018}. The grain surface chemistry includes hydrogenation reactions to form OH, H$_2$O, NH$_3$, HCN, CH$_4$, H$_2$CO, and CH$_3$OH. CO, CO$_2$, O$_2$, N$_2$, and CH$_3$CN are also synthesized on grains. We added further reactions \citep{harada2010} and updated the photodissociation cross sections \citep{heays2017}, as described by \citet{anderson2021}. The code uses a chemical kinetics approach, with each reaction/process set by a rate coefficient. Currently, our chemical evolution model includes 654 species and 7039 processes including chemical reactions in the gas and ice phases, ionization, molecular dissociation, recombination, adsorption onto grains, and desorption off of grains (thermal and nonthermal). The evolution was calculated at 100 time steps between 1 yr and 3 Myr. The code is run in 1+1D, where each disk radius is treated independently and self-shielding is accounted for in the vertical direction. 

The stellar properties and the disk gas and dust distributions remain static during the chemical evolution in our model. In reality these properties evolve during the disk's lifetime, potentially influencing the chemical evolution. See \citet{price2020} for a recent study coupling midplane chemistry with evolving disk structure and stellar properties, and see \citet{krijt2020} for a recent study coupling diffusion, grain growth, settling, and radial drift with chemical evolution. In our case, the TW Hya UV template spectrum we use reflects an older star with a lower accretion rate. Using a brighter UV spectrum early in the disk evolution would result in a warmer gas disk, higher ionization, and the concomitant effects on disk chemistry. However, coupling star and disk evolution to the chemical evolution is computationally expensive and beyond the scope of this work.

\begin{deluxetable}{lcc}
\tablecolumns{3}
\tablecaption{Initial Chemical Abundances Relative to Total H Atoms \label{table:initialabundances}}
\tablehead{\colhead{Species}  & \colhead{Inheritance } & \colhead{Reset }\\
  & (Molecular) & (Atomic)}
\startdata
H$_2$ & $5.00 \times 10^{-1}$ & $5.00 \times 10^{-1}$ \\ 
He & $1.40 \times 10^{-1}$ & $1.40 \times 10^{-1}$ \\
H$_2$O ice & $8.00 \times 10^{-5}$ & \nodata \\ 
CO & $9.92 \times 10^{-5}$ & \nodata \\ 
CO$_2$ ice & $2.24 \times 10^{-5}$ & \nodata \\ 
CH$_3$OH ice & $4.80 \times 10^{-6}$ & \nodata \\ 
NH$_3$ ice & $4.80 \times 10^{-6}$ & \nodata \\ 
CH$_4$ ice & $3.60 \times 10^{-6}$ & \nodata \\ 
N$_2$ & $3.51 \times 10^{-5}$ & \nodata \\ 
HCN & $2.00 \times 10^{-8}$ & \nodata \\ 
H$_3^+$ & $1.00 \times 10^{-8}$ & \nodata \\ 
HCO$^+$ & $9.00 \times 10^{-9}$ & \nodata \\ 
C$_2$H & $8.00 \times 10^{-9}$ & \nodata \\ 
SO & $5.00 \times 10^{-9}$ & \nodata \\ 
CS & $4.00 \times 10^{-9}$ & \nodata \\ 
O & \nodata & $2.29 \times 10^{-4}$ \\
C & \nodata & $1.30 \times 10^{-4}$ \\
N & \nodata & $7.50 \times 10^{-5}$ \\
S & \nodata & $9.00 \times 10^{-9}$ \\
Si$^+$ & $1.00 \times 10^{-11}$ & $1.00 \times 10^{-11}$ \\ 
Mg$^+$ & $1.00 \times 10^{-11}$ & $1.00 \times 10^{-11}$ \\ 
Fe$^+$ & $1.00 \times 10^{-11}$ & $1.00 \times 10^{-11}$  
\enddata
\end{deluxetable}

\subsection{Ice mass and opacity}
\label{sec:icemassopacity}
We use the results of the chemical evolution model to set up a radiative transfer model using RADMC-3D \citep{dullemond2012} to simulate observations of the ice features from the disk. While the chemical model yields the abundance distribution of many species at numerous time steps, we extracted the abundances of six ice species (H$_2$O, CO, CO$_2$ CH$_3$OH, NH$_3$, and CH$_4$) at four time steps (0.25, 0.5, 1, and 2 Myr) to use in the radiative transfer model. These six species were chosen because they are among the most abundant ices with observable spectral signatures.

To set up the radiative transfer model, we need the dust-plus-ice mass density and the composition (volume fraction of each species) at each location in the disk. We compute the mass density of each ice species from
\begin{equation}
\label{eq:rhoice}
\rho_\text{ice} = \rho_g\frac{x_\text{ice} M_\text{ice}}{x_g M_g},   
\end{equation}
where $x_\text{ice}$ is the abundance of the ice species (from the chemical model) and $M_\text{ice}$ is its molar mass. The gas, primarily composed of H$_2$ and He, has an abundance of $x_g$ = 0.64 (relative to total H atoms) and a molar mass of $M_g$ = 2.44.

Because we model the small- and large-grain populations separately, we apportion the ice mass density between them. We do so assuming that the local ratio of ice mass between the two populations is proportional to local ratio of dust surface area, as expected for ices growing on the dust grains. With this assumption, the fraction of ice mass density apportioned to the large grains is
\begin{equation}
\label{eq:icefrac}
f_{\text{ice,l}} = \frac{f_{\text{dust,l}}}{\sqrt{a_\text{max,l}/{a_\text{max,s}}} (1-f_{\text{dust,l}}) + f_{\text{dust,l}}}    
\end{equation}
with the rest in the small grains. In Equation \ref{eq:icefrac}, $f_{\text{dust,l}}$ is the local fraction of the dust mass density in the large-grain population, and $\sqrt{a_\text{max,l}/{a_\text{max,s}}} = \sqrt{1000}$ is the ratio of surface area to mass of the small-grain population over that of the large-grain population with $q$ = 3.5. 

This yields the mass density of each ice species associated with each dust population at each location in the disk. However, we mix the ices and dust according to their volume fractions, not mass fractions. Thus, we compute the volume density (volume of material per volume of space) $V = \rho/\rho_m$ for the dust and ice species throughout the disk, where $\rho_m$ is the material density of each species listed in Table \ref{table:iceproperties}. The volume fraction of each species ($i$) is then $V_i / \sum V_i$.

\begin{deluxetable}{lcc}
\tablewidth{0pt}
\tablecolumns{3}
\tablecaption{Ice and Dust Properties \label{table:iceproperties}}
\tablehead{\colhead{Species}  & \colhead{$\rho_m$ (g cm$^{-3}$)} & \colhead{Optical Constants Ref.}}
\startdata
H$_2$O & 0.87 & \cite{warren2008} \\ 
CO & 0.80 & \citet{palumbo2006} and \\
& &  \citet{giuliano2019} \\
CO$_2$ & 1.11 & \citet{baratta1998} \\ 
CH$_3$OH & 1.01 & \citet{hudgins1993} \\ 
NH$_3$ & 0.74 & \citet{trotta1996}\tablenotemark{a} \\ 
CH$_4$ & 0.45 & \cite{HudsonWebsite} \\ 
Silicates & 3.3 & \citet{draine2003} \\
Graphite & 2.2 & \citet{draine2003} 
\enddata
\tablecomments{Material density values ($\rho_m$) for the ice species are from Table 2 of \citet{bouilloud2015}.}
\tablenotetext{a}{These optical constants were accessed from the SSHADE database \citep{SSHADE}.}
\end{deluxetable}

\begin{figure*}
\epsscale{1.15}
\plotone{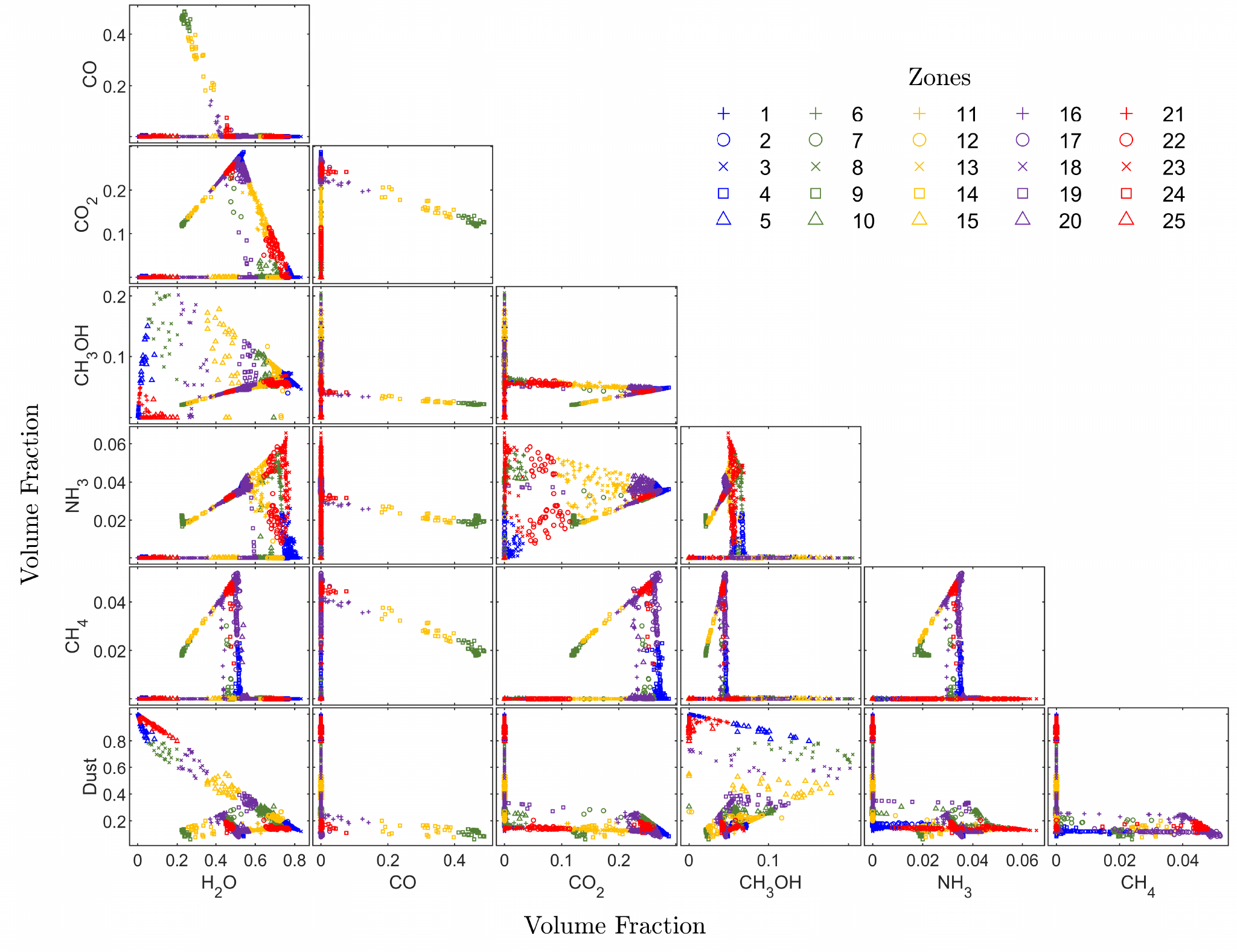}
\caption{Results of the \textit{k}-means cluster-finding algorithm to identify zones of similar composition. Each point represents a cell in the chemical model with composition defined by its volume fraction of six ice species and dust. Each panel shows a 2D projection of this 7D parameter space. The color and symbol of the points indicate the zone to which they are assigned by the \textit{k}-means algorithm. The specific numbering of the zones is arbitrary. Shown here are the results for the small-grain population of the Inheritance chemical model after 0.25 Myr of evolution.}
\label{fig:clustering}
\end{figure*}

\begin{figure*}
\epsscale{1.17}
\plotone{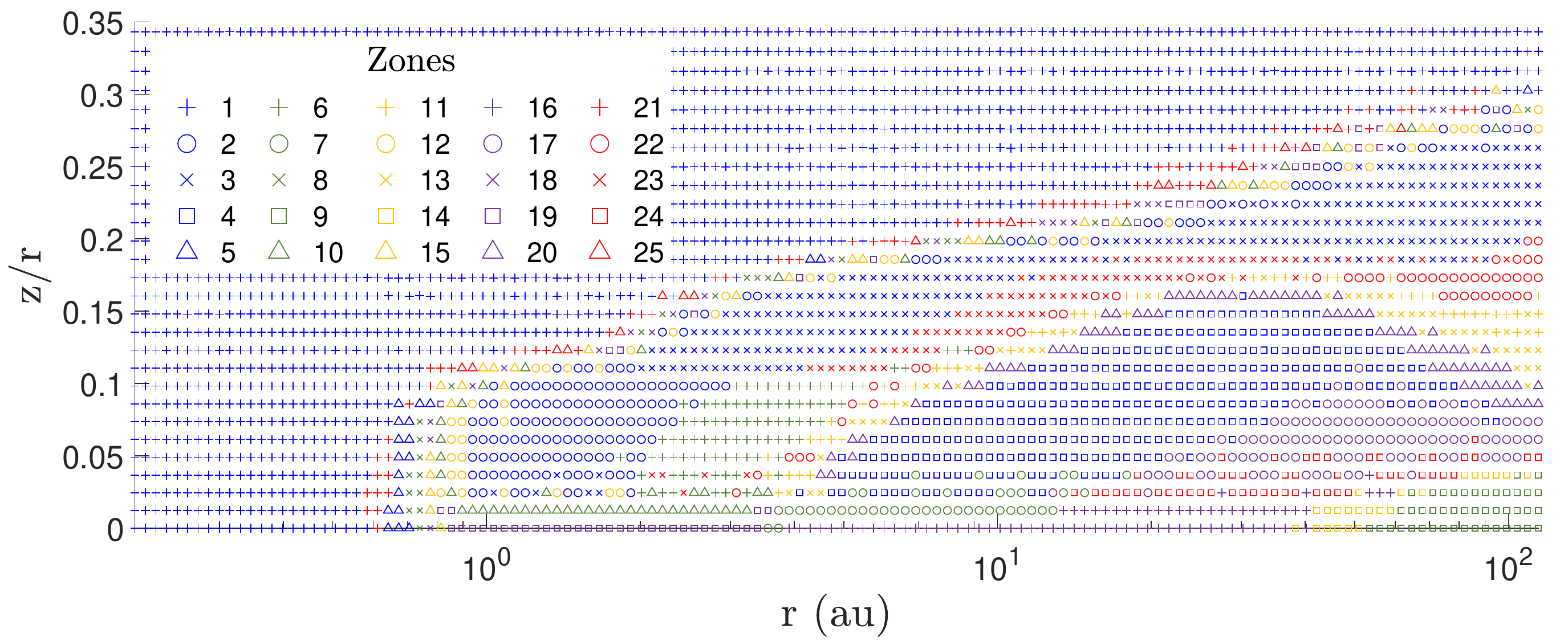}
\caption{Location in the disk of compositionally similar zones. Shown here is the same model shown in Figure \ref{fig:clustering} (small grains, Inheritance chemistry, 0.25 Myr). Each point represents a cell in the chemical model with color and symbol identifying the zone to which it is assigned by the \textit{k}-means algorithm. The plot is truncated at $z/r$ = 0.35 although the model includes cells up to $z/r$ = 0.7; the upper cells all belong to zone 1.}
\label{fig:zonelocations}
\end{figure*}

Each point in the chemical model has a unique mixture of dust and ices, so ideally we would use a unique opacity for each spatial cell in the radiative transfer model. However, RADMC-3D requires, for each opacity, the density to be specified in \textit{every} cell. This ideal setup would thus require a number of opacity spectra and density profiles equal to the number of spatial cells, with each density profile having zero values in every cell except the one cell of the corresponding opacity.  Implementing this in a brute force way would be highly impractical and computationally expensive. We overcome this issue by grouping cells from the chemical model into ``zones" with similar volume fractions of dust and ice species, and thus similar opacities. To find the zones, we use a \textit{k}-means cluster-finding algorithm.\footnote{We implement this with the \texttt{kmeans} function in MATLAB, using a cityblock distance metric and 100 replications with new initial cluster centroid positions to find the optimal convergence.} After some experimentation, we found that using 25 zones for each dust population could sample the full range of grain compositions while allowing the radiative transfer simulation to be computationally tractable. We use the median volume fraction of each species for the representative volume fraction assigned to all cells in the zone. The results of our zone-finding procedure are illustrated in Figures \ref{fig:clustering} and \ref{fig:zonelocations}. For the large-grain population, one zone always identified cells from the upper parts of the disk containing no ice and no large grains; these cells contained no solid mass so this zone is ignored for the remainder of the modeling.    

We gathered laboratory-measured optical constants from the literature (see Table \ref{table:iceproperties} for references) for the dust (silicates and graphite) and ices. We used a single set of optical constants for each species and thus did not account for how the strength and precise spectral location of ice features vary with temperature. Each set of optical constants was interpolated onto a common grid of 683 wavelength points from 0.1 to 10$^4$ $\micron$ with dense sampling in the near-to-mid-IR, especially where the ices exhibit their characteristic spectral features, and coarser sampling outside the ice features. While silicates, graphite, and water ice have optical constants measured over a wide wavelength range, those of the other ices are limited primarily to the IR. The imaginary part of these optical constants is close to zero at the edges of the measured wavelength ranges, while the real part has a constant value, so we extrapolated the optical constants at these values to wavelengths where they are not measured. We then used the Bruggeman mixing rule to combine the constants into a representative set of optical constants for each zone. By using the Bruggeman mixing rule, we assume the grains are an intimate mixture of the constituent species.

\begin{figure}
\epsscale{1.15}
\plotone{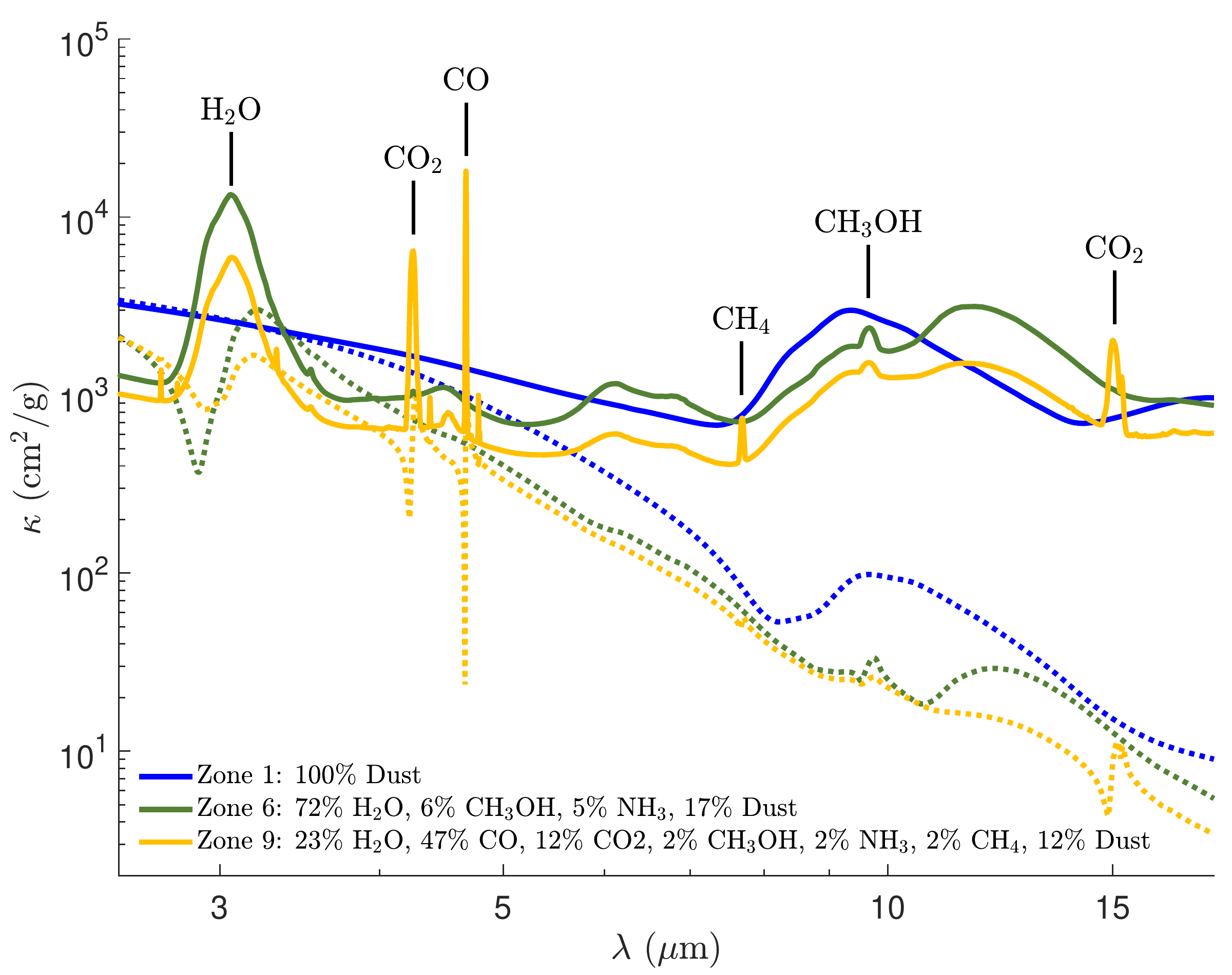}
\caption{Example absorption (solid lines) and scattering (dotted lines) opacity spectra for dust and ice mixtures from three example zones. Zone 1 is ice-free, zone 6 is H$_2$O ice-rich, and zone 9 is CO and CO$_2$ ice-rich. These zones are for the small-grain population of the Inheritance chemical model after 0.25 Myr of evolution (as shown in Figures \ref{fig:clustering} and \ref{fig:zonelocations}).}
\label{fig:OpacitySpectra}
\end{figure}

We computed the absorption and scattering opacity spectra from the optical constants using Mie theory with a code included in the RADMC-3D software package based on the algorithm by \citet{bohren1983}. The opacity was computed at 100 grain sizes between $a_\text{min}$ and $a_\text{max}$ and averaged according to the grain size distribution. Example infrared opacity spectra of three small-grain zones are plotted in Figure \ref{fig:OpacitySpectra}. Zone 1 illustrates the opacity spectrum of ice-free dust. Zone 6 illustrates an H$_2$O ice-rich mixture. It also has the highest volume fraction of NH$_3$ ice of any zone in the model, yet no strong features of this ice are visible in the opacity spectrum. This suggests, even before running radiative transfer simulations, that NH$_3$ ice  will be difficult to observe. In contrast, CH$_3$OH ice, with a similar volume fraction, does show a clear feature in the absorption opacity. Zone 9 is rich in CO and CO$_2$ ice, and their spectral features are prominent. CH$_4$ and  CH$_3$OH ice are present at small fractions in this zone, and their features are evident. For all zones, the scattering and absorption opacities are of comparable amplitude at shorter wavelengths, while the scattering opacity becomes much weaker at longer wavelengths, as expected for small grains.  

The Mie theory code also computes the full scattering matrix at 181 angles equally spaced from 0$^\circ$ to 180$^\circ$, allowing for a realistic treatment of the scattering during the radiative transfer. We truncate the phase function within 3$^\circ$ of forward-scattering because it rises sharply toward the peak and is difficult to sample properly. This has minimal impact on the radiative transfer. 

\subsection{Simulating Disk Observations}

The chemical evolution model used a cylindrical coordinate grid. However, cylindrical coordinates are not supported in RADMC-3D, so for the radiative transfer simulation we used a spherical coordinate grid with sufficient cell density to sample the chemical model. The grid has 180 cells in the radial dimension, 120 in the polar dimension, and 120 cells in the azimuthal dimension. We interpolate the properties of the chemical model (ice mass and opacity) to the RADMC-3D spherical grid.

We set up the disk model using the 49 opacities (25 for small grains, 24 for large grains) computed as described in Section \ref{sec:icemassopacity}. For each opacity, we populate the cells of the corresponding zone with the total solid mass: dust mass (Equation \ref{eq:smalldensity} or \ref{eq:largedensity}) augmented with the total ice mass of the six species we considered, apportioned to each dust population (Equations \ref{eq:rhoice} and \ref{eq:icefrac}).     

We recompute the temperature throughout the disk with RADMC-3D using 10$^7$ photons for the radiative transfer. This differs somewhat from the initial calculation of temperature because of the additional solid mass and different opacities due to the inclusion of ices. We note that the opacities in the UV and optical (where the stellar radiation field efficiently heats the solids) differ by only a factor of two between zones, and thus we do not expect this to have a major impact on the disk temperature. We accelerate the radiative transfer in the densest parts of the disk with a modified random walk procedure.

We then simulate images of the disk with RADMC-3D at each of the 683 wavelengths. The image size was 280 $\times$ 280 au sampled with 400 $\times$ 400 pixels. We sum the flux in each image to make a model spectrum and scale the amplitude for a disk residing at a distance of 100 pc. The flux from edge-on disks in the near-to-mid-IR is dominated by photons emitted by the star and hot inner disk then scattered at the disk surface before passing through the colder outer disk. Proper sampling of this effect requires using a large number of photons. We adopt 10$^{10}$ photons for images at 2.5--8 $\micron$ and 10$^8$ photons outside of this wavelength range. The models using 10$^{10}$ photons require a significant amount of computing time ($\sim$50--75 hr per wavelength). We also produce a set of images with scattering turned off in the radiative transfer to assess the importance of scattering in the primary models using 10$^8$ photons at all wavelengths.

To explore the effects of disk inclination on the observable ice features, we generated images of the Inheritance model at five inclinations (0$^\circ$, 75$^\circ$, 80$^\circ$, 85$^\circ$, and 90$^\circ$) at one time step (0.25 Myr), and to explore the effect of disk evolution on the features we made images at three additional time steps (0.5, 1, and 2 Myr) at one inclination (90$^\circ$). To test the effect of initial abundances, we generated model images of the Reset model at all four time steps at 90$^\circ$ inclination. Tabulated disk-integrated spectra for these models are available in electronic form (Table \ref{table:simulations}). 

\begin{deluxetable}{lcc}
\tablecolumns{3}
\tablecaption{Summary of Simulated Observations \label{table:simulations}}
\tablehead{\colhead{Chemical Model}  & \colhead{$i$} & \colhead{$t$} \\ \colhead{} & \colhead {(deg)} & \colhead{(Myr)}}
\startdata
Inheritance & 0 & 0.25 \\ 
Inheritance & 75 & 0.25 \\
Inheritance & 80 & 0.25 \\
Inheritance & 85 & 0.25 \\
Inheritance & 90 & 0.25 \\
Inheritance & 90 & 0.5 \\
Inheritance & 90 & 1.0 \\
Inheritance & 90 & 2.0 \\
Reset & 90 & 0.25 \\
Reset & 90 & 0.5 \\
Reset & 90 & 1.0 \\
Reset & 90 & 2.0
\enddata
\tablecomments{The simulated disk-integrated spectra for these 12 models are provided in a machine-readable format.}
\end{deluxetable}

\section{Results and Analysis}
\label{sec:results}

\subsection{Ice Abundances}
\label{sec:abundances}

\begin{figure*}
\epsscale{1.17}
\plotone{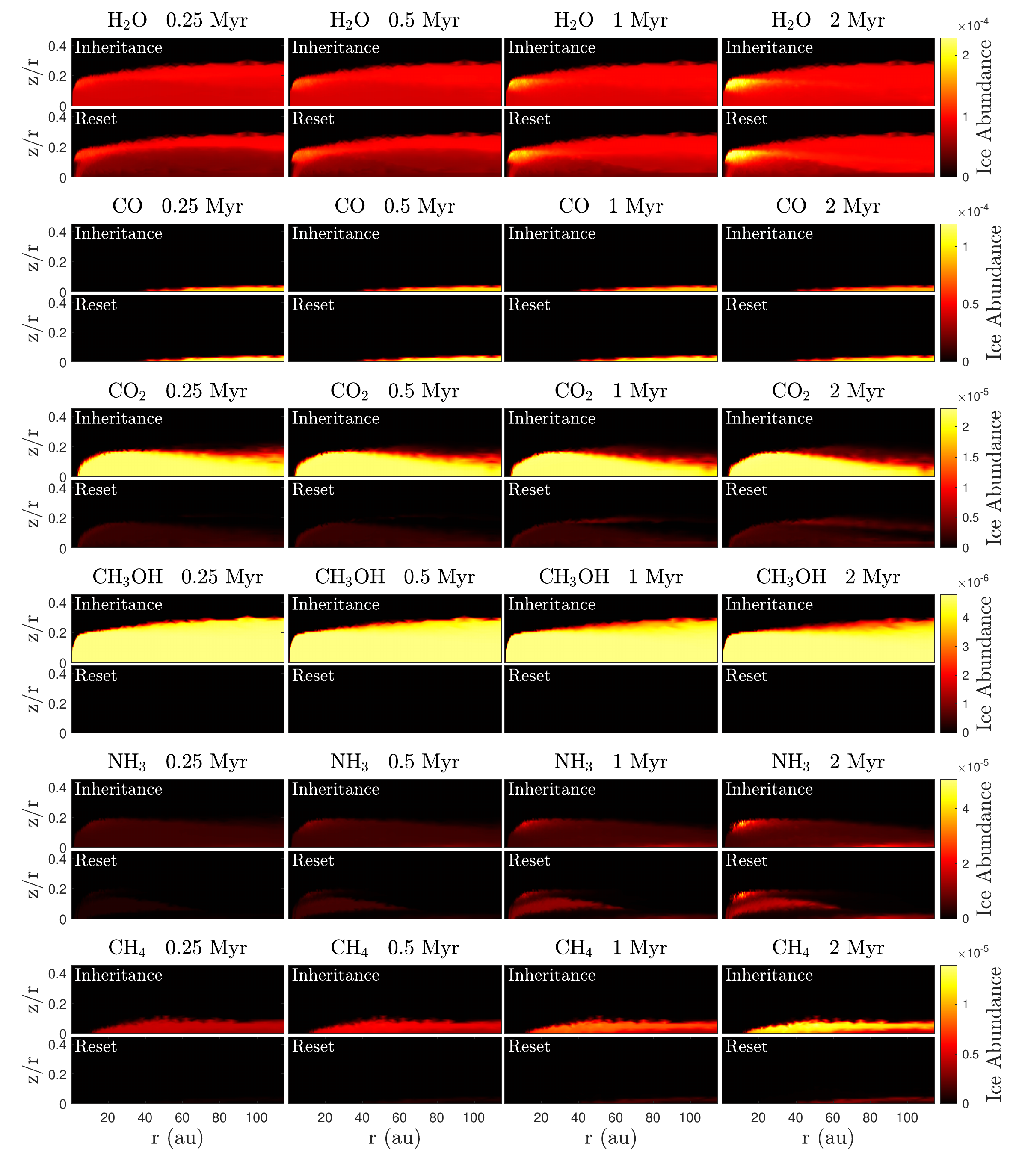}
\caption{Abundance distribution (relative to the total number of H atoms) of the six ice species of interest at four time steps. For each species, the top row is the Inheritance chemical model and the lower row is the Reset model. Note the linear scaling of the color bar, which is more relevant for comparing the observable total ice content between models and over time.}
\label{fig:abundances}
\end{figure*}

Figure \ref{fig:abundances} shows the ice abundances at 0.25, 0.5, 1, and 2 Myr. H$_2$O is the most abundant of the six ices. It is broadly distributed radially and vertically in the disk, although more concentrated in the surface layers in the Reset model, in the irradiated part of the warm molecular layer. In both models, the H$_2$O abundance grows with time, most noticeably in the inner and upper region of the disk.

CO ice, due to its high volatility, is confined to the disk midplane, where the temperatures are below the sublimation temperature of $\sim$17 K. There is little difference in the abundance between the two chemical models, and in both, the abundance decreases by $\sim$12\% over time.

CO$_2$ ice is distributed broadly throughout the disk in both models, albeit at higher overall abundance in the Inheritance model. The primary difference is in the upper and outer region, where the abundance decreases with time in the Inheritance model but increases in the Reset model, reaching similar abundances by 2 Myr. 

CH$_3$OH ice shows the starkest difference between the two chemical models. In the Inheritance model, it is distributed from the midplane to a similar vertical height to the H$_2$O ice. The abundance in the uppermost layers decreases somewhat over time. CH$_3$OH ice is several orders of magnitude less abundant in the Reset case, revealing that disk chemistry alone is not efficient in producing CH$_3$OH, even though both gas and grain surface pathways are included in our network. Perhaps CH$_3$OH formation could be enhanced under elevated cosmic ray conditions or high stellar X-ray luminosities \citep{bosman2018,schwarz2018,krijt2020}.

NH$_3$ ice is somewhat more abundant and has a broader distribution in the Inheritance model than in the Reset model. The ice increases with time in distinct locations: the upper inner edge of the distribution in both models, the outer midplane in both models, and intermediate heights in the inner half of the disk in the Reset model only. 

In the Inheritance model, CH$_4$ ice is concentrated at low disk heights but peaks somewhat above the midplane in the outer disk. In the Reset model, CH$_4$ ice exists only at very low heights and large radii. In both models the abundance increases with time.

\subsection{Effect of Disk Inclination}
\label{sec:inclination}

Observing the near-to-mid-IR ice absorption features requires a source of background illumination, which in the case of protoplanetary disks arises from the central star and warm inner disk. For low-inclination systems, the flux is dominated by direct emission from these regions with no intervening ice. We find no ice absorption features in the disk-integrated spectra of our face-on model, as expected, and focus this section on the more edge-on cases. Scattered-light features from a face-on disk are discussed in Section \ref{sec:spatial}.

\begin{figure}
\epsscale{1.15}
\plotone{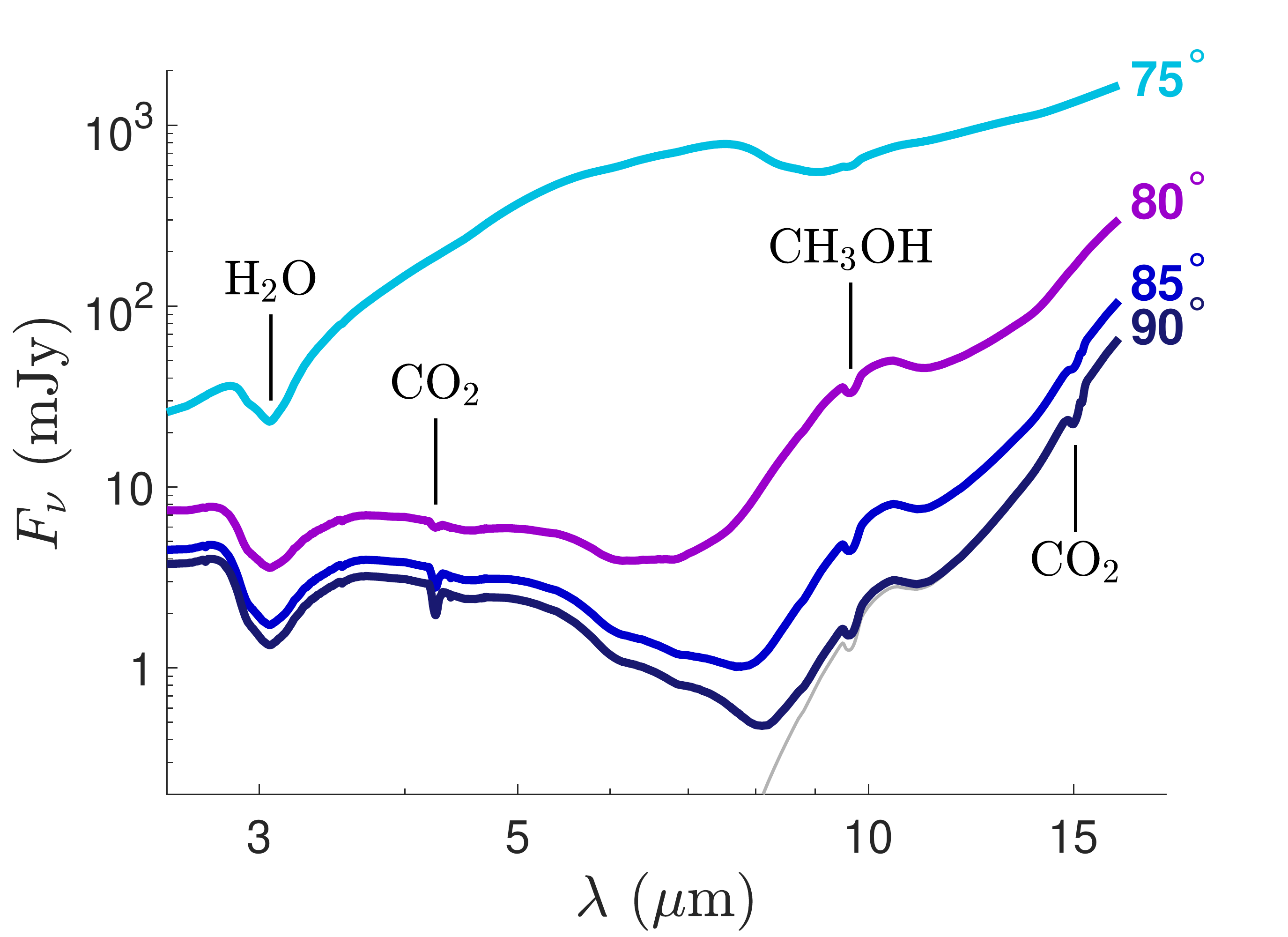}
\caption{Model near-to-mid-IR spectra at four disk inclinations: 75$^\circ$, 80$^\circ$, 85$^\circ$, and 90$^\circ$. The prominent spectral features of H$_2$O, CO$_2$ and CH$_3$OH are labeled. The thin gray line is the model spectrum at 90$^\circ$ with scattering turned off in the radiative transfer. These spectra are from the Inheritance chemical model after 0.25 Myr.}
\label{fig:InclinationEffect}
\end{figure}

Figure \ref{fig:InclinationEffect} shows the effect of inclination on the disk spectra and observable near-to-mid-IR ice features. Higher inclination allows the outer disk to extinct more of the background radiation, resulting in fainter emission at all wavelengths. This geometry also leads to more radiation passing through ice-rich regions, enhancing the strength of the ice features relative to the continuum. At 75$^\circ$, H$_2$O ice is readily detectable and the CH$_3$OH feature is barely discernible. At 80$^\circ$, CH$_3$OH becomes detectable and the CO$_2$ feature at 4.25 $\micron$ is evident, albeit weak. At 85$^\circ$ and 90$^\circ$, CO$_2$ is also detectable at 15 $\micron$, and the features become more prominent. This trend reflects the vertical distribution of ices seen in Figure \ref{fig:abundances}: H$_2$O and CH$_3$OH ices exist in higher layers in the disk than CO$_2$.

The gray line in Figure \ref{fig:InclinationEffect} is the model at 90$^\circ$ inclination with scattering turned off in the radiative transfer. This model shows that at $\lambda \lesssim 8$ $\micron$ the flux is dominated by photons that were scattered by dust near the disk surface, effectively magnifying the size of the background radiation source and allowing the photons to leave the system above (and below) the dense disk midplane. At $\lambda \gtrsim 10$ $\micron$ the gray curve converges with the full model, indicating that scattering does not contribute significantly, and the background radiation source is due to direct thermal emission. As also discussed by \citet{pontoppidan2005}, the shape and size of the background source change with wavelength. The 4.25 $\micron$ CO$_2$ feature is illuminated by scattered light while the 15 $\micron$ CO$_2$ feature is illuminated by thermal emission, so these features likely do not probe identical CO$_2$ ice reservoirs. These effects underscore the importance of interpreting disk ice observations with radiative transfer simulations that properly model scattering.   

What of the three ice species with no features seen in these spectra? CO is quite abundant but confined to the optically thick disk midplane, which does not contribute significantly to the total disk flux. NH$_3$ and CH$_4$ are less abundant than the detectable species. For the Inheritance model at 0.25 Myr, the most NH$_3$-rich and CH$_4$-rich grains each contain only $\approx$5\% of that species by volume in the small-grain population (Figure \ref{fig:clustering}). For reference, the most CH$_3$OH-, CO$_2$-, CO-, and H$_2$O-rich grains are composed of $>$15\%, $>$25\%, $>$45\%, and $\approx$80\% of these species, respectively. In addition to their low abundances, CH$_4$ is absent near the disk surface while NH$_3$ ice has intrinsically weak spectral features as (not) seen in Figure \ref{fig:OpacitySpectra}.

\begin{figure}
\epsscale{1.15}
\plotone{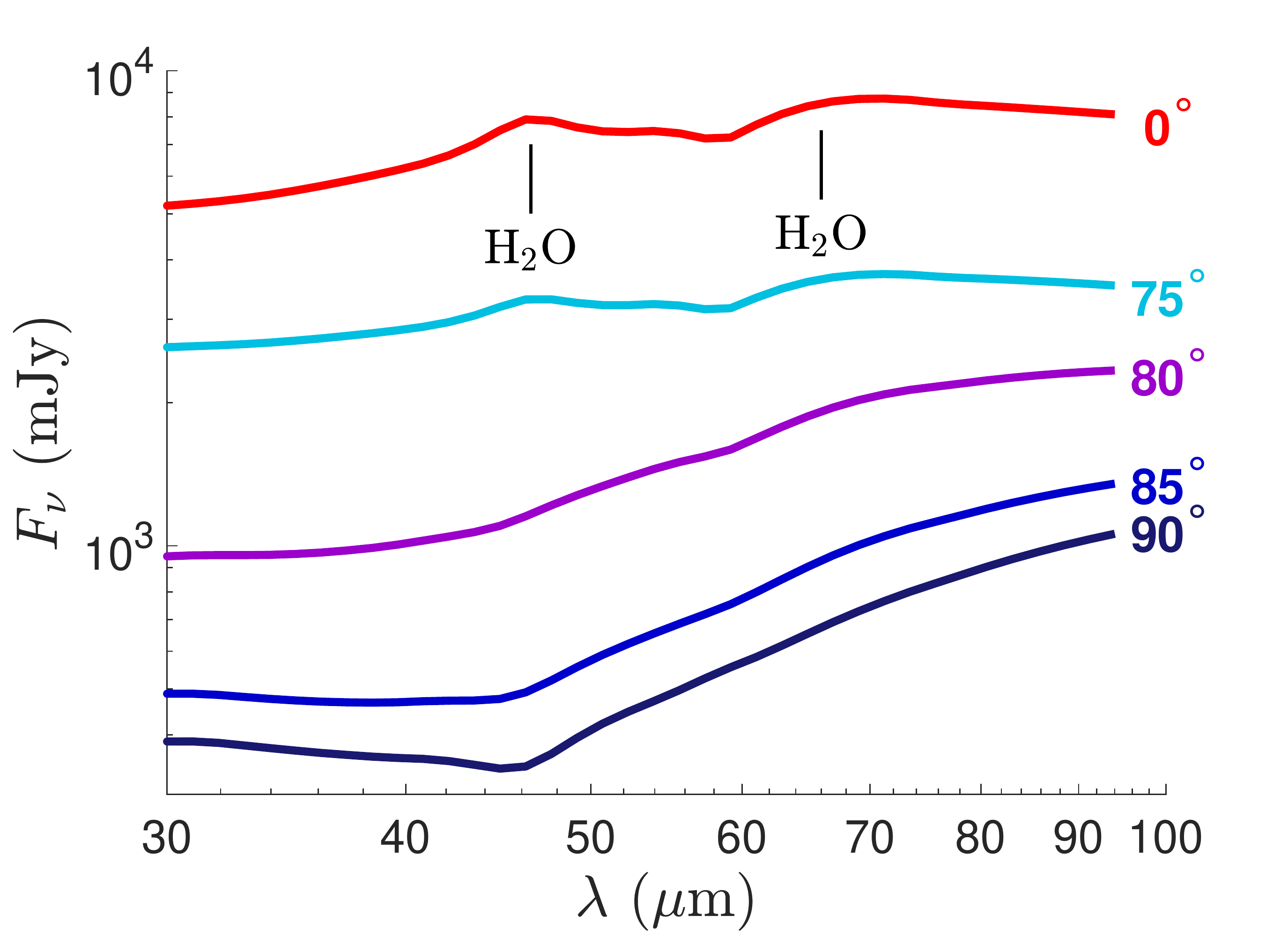}
\caption{Model far-IR spectra at five disk inclinations: 0$^\circ$, 75$^\circ$, 80$^\circ$, 85$^\circ$, and 90$^{\circ}$. The H$_2$O ice emission features are seen prominently for a face-on disk but not for disks viewed close to edge-on. These spectra are from the Inheritance chemical model after 0.25 Myr.}
\label{fig:FIR}
\end{figure}

We find that the far-IR emission features from water ice have distinctly different behavior from the near-to-mid-IR features. Figure \ref{fig:FIR} illustrates this effect. We find that the continuum is brighter and the ice features more prominent at lower (closer to face-on) inclinations. Thus, while edge-on disks are the preferred targets to study disk-integrated near-to-mid-IR absorption features, face-on disks are preferred when studying the far-IR emission features.

We used optical constants from crystalline (rather than amorphous) water ice in our model \citep{warren2008}. Crystalline water ice gives rise to more prominent far-IR features than does amorphous ice, making our simulations an optimistic case for observable features. The degree to which water ice in disks is truly crystalline is unknown. Crystalline water is seen in outer solar system objects \citep{jewitt2004}, and models of disks with detected far-IR features reveal that most of the water ice is crystalline, despite the ice having temperatures too low for it to thermally crystallize in situ \citep{mcclure2015,min2016}. This crystallinity could arise from the transport of material from warmer regions, from collisions, or from past transient heating events (i.e. accretion outbursts). Whether the nondetection of these features in other disks is due to amorphous ice or other effects---e.g., large grain sizes \citep{kamp2018}, vertically settled icy grains, or small disk radii \citep{mcclure2015}---is not known.

\subsection{Spatially Resolved Ice Features}
\label{sec:spatial}

\begin{figure}
\epsscale{1.15}
\plotone{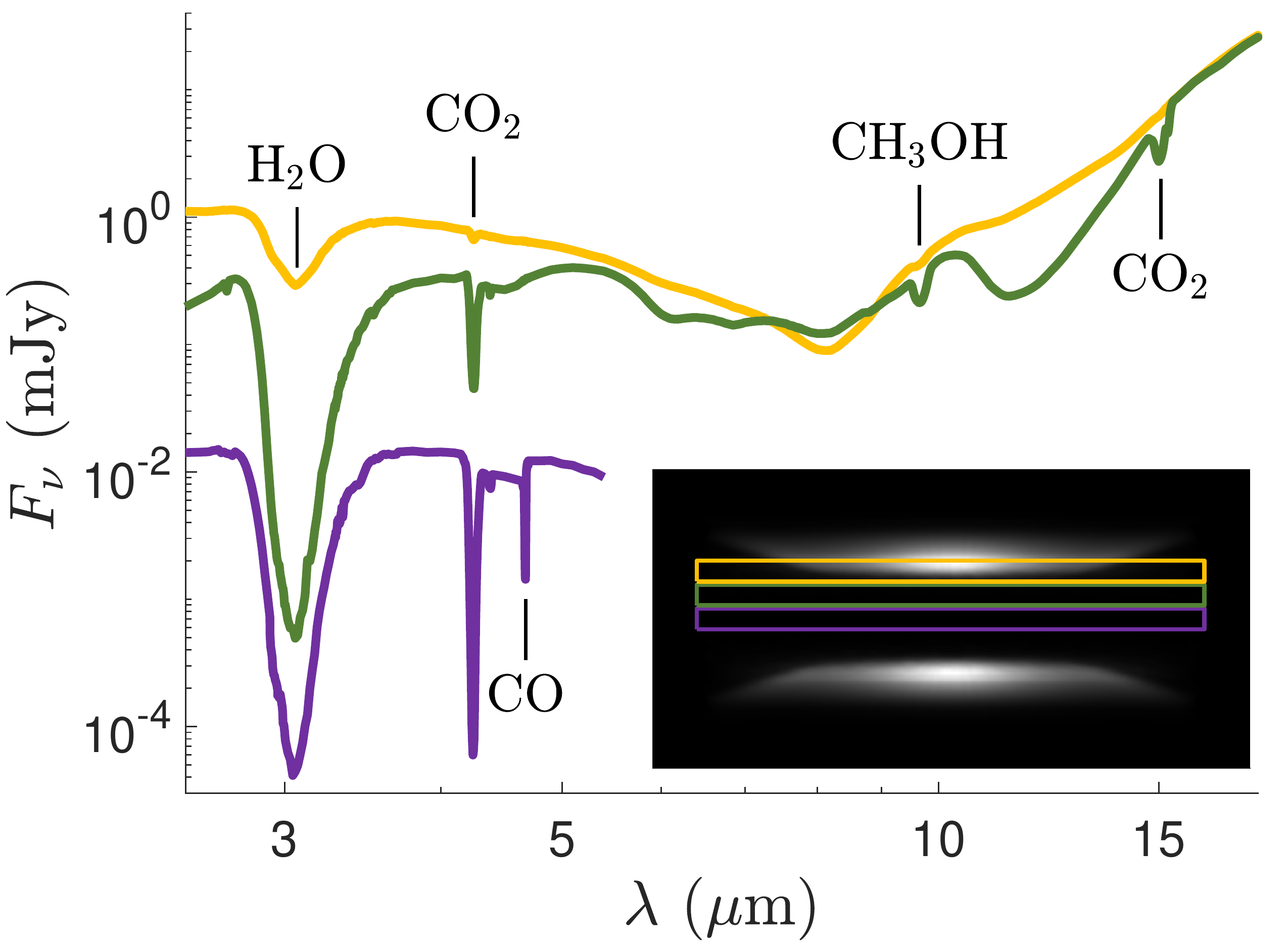}
\caption{Spectra extracted from three apertures of a 90$^\circ$ inclination disk model probing different vertical heights. The prominent spectral features of H$_2$O, CO, CO$_2$ and CH$_3$OH are labeled. The inset shows the model image at 0.9 $\micron$ overlaid with the locations of the apertures. Each aperture has a width of 9.8 au. We truncate the purple spectrum from the disk midplane at 5.5 $\micron$, beyond which it suffers from strong numerical noise.}
\label{fig:VerticalApertures}
\end{figure}

\begin{figure}
\epsscale{1.15}
\plotone{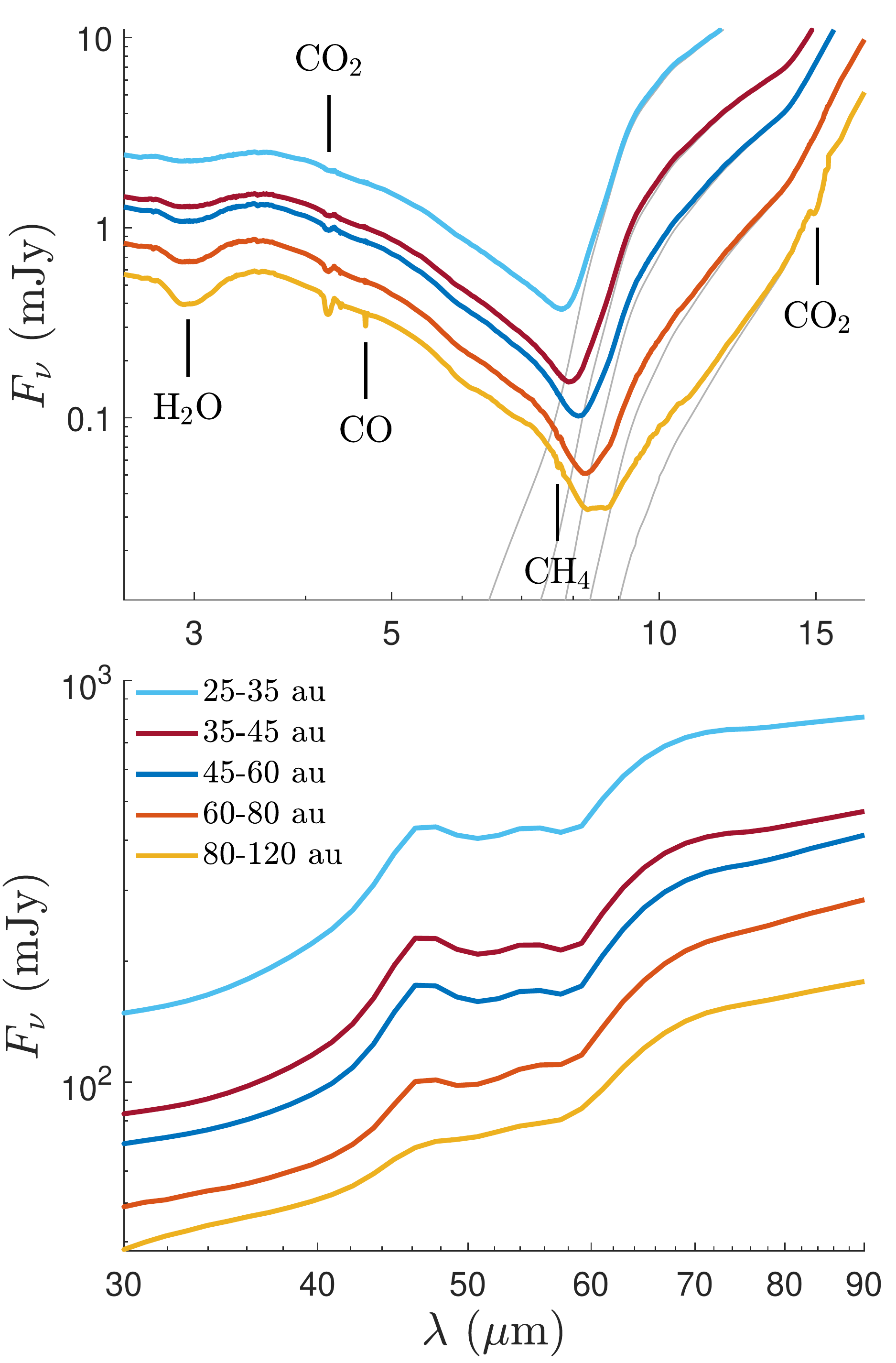}
\caption{Spectra extracted from annular apertures of a face-on model disk. The top panel shows the near-to-mid-IR region with ice features labeled; the bottom panel shows the far-IR region around the H$_2$O emission features. The thin gray lines are the model spectra with scattering turned off in the radiative transfer.}
\label{fig:SpatialFaceOn}
\end{figure}

Our models consist of images at each wavelength, allowing us to predict the results of spatially resolved disk spectra, as might be obtained from IFU spectroscopy from the NIRSpec and MIRI JWST instruments. Figure \ref{fig:VerticalApertures} presents spectra extracted from a 90$^\circ$ inclination model in three rectangular apertures at different disk heights. The inset shows the model disk image at 0.9 $\micron$---exhibiting the bi-lobed emission pattern on either side of the dark midplane characteristic of edge-on disk images---overlaid with the aperture locations. Each aperture has a width of 9.8 au (14 pixels in the model image). We chose this width so that at a distance of 100 pc the angular scale (0$\farcs$098) corresponds to the size of the JWST/NIRSpec IFU spaxels (0$\farcs$1). 

Emission from higher in the disk (yellow) exhibits a clear H$_2$O feature and weak CO$_2$ and CH$_3$OH features. Deeper in the disk (green), the features from these three species become significantly more prominent. At the disk midplane (purple), a significant CO ice feature is present, although the overall faintness of the emission in this aperture would make it difficult to observe in practice. The purple spectrum suffered from strong numerical noise beyond 5.5 $\micron$, so we do not plot that region. Overall, this trend in observable features with disk height tracks the vertical location of the ices shown in Figure \ref{fig:abundances}. Such observations will be useful to test predictions of the chemical model in more detail than is possible with only disk-integrated observations.

We next investigate spatially resolved ice features in our face-on disk model. The disk-integrated near-to-mid-IR flux is dominated by emission from the star and inner disk, so no ice features are evident for a spatially unresolved spectrum. However, spatially isolating the scattering-dominated outer parts of the disk does reveal ice features. The top panel of Figure \ref{fig:SpatialFaceOn} presents near-to-mid-IR spectra extracted from annular apertures at a range of disk radii. The gray lines are again radiative transfer models run without scattering, shown to illustrate the wavelength range where scattered light dominates. H$_2$O and CO$_2$ appear only weakly at 25--35 au but become more prominent with increasing disk radius. Spectra from the outermost radii (80--120 au) also exhibit the CO feature and a weak, but present, CH$_4$ feature. Thus, spatially resolved spectra can be a powerful tool to study ices in all disks, not just those that are edge-on. In practice, isolating flux from the outer part of a face-on disk is complicated by the glare from the star and inner disk and often requires coronagraphy. 

The features seen in scattered light are shifted to slightly shorter wavelengths than those seen from absorption in the edge-on disk spectra. This is because the minimum in the scattering opacity occurs at a shorter wavelength than the maximum in the absorption opacity for each feature (see Figure \ref{fig:OpacitySpectra}). This may be useful in interpreting observations where the origin of the feature is unclear, e.g., in intermediate-inclination systems where the central star is occulted but significant flux reaches the observer after being scattered from the surface of the outer disk without then passing through an ice-rich part of the disk.

The bottom panel of Figure \ref{fig:SpatialFaceOn} presents the far-IR region of the spectra from the same annular apertures. The H$_2$O emission features are present with similar strengths relative to the continuum at all radii except the outermost annulus, where the features are weaker.

\subsection{Initial Abundances and Evolution}
\label{sec:intialandtime}

\begin{figure}
\epsscale{1.16}
\plotone{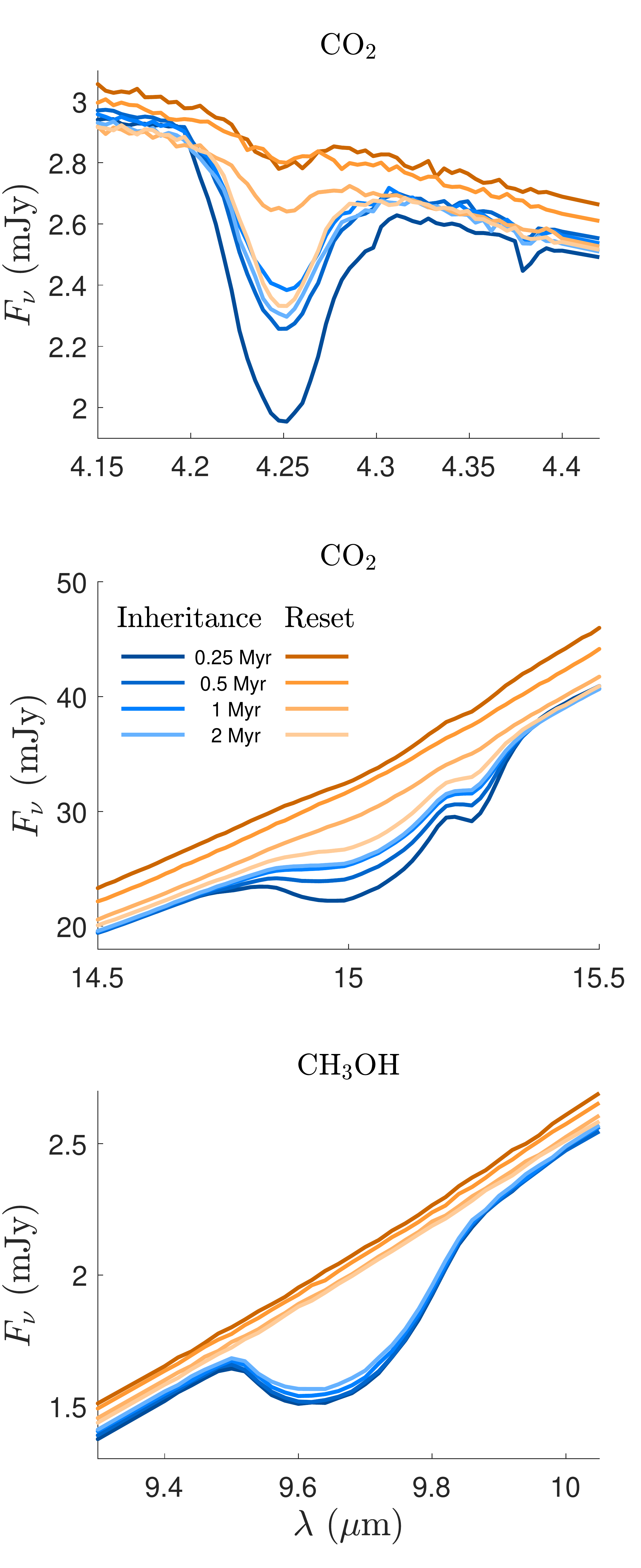}
\caption{Features from CO$_2$ ice (top and middle panels) and CH$_3$OH ice (bottom panel) for the Inheritance (blue) and Reset (orange) chemical models at four time steps (evolving from dark to light). Small-amplitude variations in the spectra are due to noise in the radiative transfer simulations.}
\label{fig:SpecvsTimevsChem}
\end{figure}

Can observations of ice discern between models that include inherited ice and those that have been fully reset? And can the chemical evolution of ice over time imprint observable signatures?  To investigate these questions, we examined each prominent ice feature seen in our 90$^\circ$ inclination disk-integrated model for variation in time and between the chemical Inheritance and Reset models. The H$_2$O 3 $\micron$ and far-IR features showed no variation with either of these properties.

CO$_2$ showed the strongest time variation, with the strength of the features in the Inheritance model decreasing between 0.25 and 2 Myr while the strength of the features in the Reset model increased (top and middle panels of Figure \ref{fig:SpecvsTimevsChem}). This change is consistent with the evolution of CO$_2$ ice abundance in the upper and outer edge of its extent (Figure \ref{fig:abundances}). Thus, the spectral features are especially sensitive to this region of the disk, in large part because it lies above the midplane of high optical depth.

CH$_3$OH shows the greatest difference between the Inheritance and Reset models, exhibiting a significant feature in the former at all time steps and no detectable feature in the latter (bottom panel of Figure \ref{fig:SpecvsTimevsChem}). This stark difference in feature strength is expected considering the drastic difference in CH$_3$OH abundance between the two chemical models (Figure \ref{fig:abundances}). 
These results bolster support for observing multiple ice features in a sample of disks with similar properties (stellar type, mass, size, etc.). Searching first for the presence or absence of CH$_3$OH ice would indicate whether the disks' initial abundances were inherited from the protostellar phase or not. Then, a comparison of the strength of their CO$_2$ features relative to their H$_2$O features could reveal whether the disks are at similar or different stages in their chemical evolution.

\subsection{Detectability with JWST}
\label{sec:JWST}

Our simulated disk-integrated spectra clearly show ice features when viewed by eye, but can they be detected in practice with the observing capabilities of JWST? We tested this using the JWST Exposure Time Calculator (version 1.6.1) for our 90$^\circ$ inclination model. We first checked that the normalization of the simulated spectra agreed with measured IR photometry of nearby edge-on disks \citep{evans2009}, and we found good agreement. The faintest 4.25 $\micron$ CO$_2$ feature (Reset model at 0.25 Myr) has a depth of 5\% of the continuum flux density. With the NIRSpec IFU G395M/F290LP configuration, detecting this feature at 3$\sigma$ confidence requires 6 minutes of exposure time while 10$\sigma$ requires 14 minutes. The 15 $\micron$ CO$_2$ feature (Reset model at 2 Myr) has a depth of 6\% and can be detected with the MIRI MRS channel 3 grating B at 3$\sigma$ (10$\sigma$) confidence in 4 (14) minutes. The Inheritance model CH$_3$OH feature, where the continuum is fainter, has a depth of 19\% and can be detected with the MRS channel 2 grating B at 3$\sigma$ (10$\sigma$) confidence in 13 (60) minutes. Note that MRS grating B observations are acquired for all four channels simultaneously, but acquiring a full mid-IR spectrum also requires observations with gratings A and C. The spectral resolution of these instruments is greater than required to resolve the relatively broad ice features, so the sensitivity can be enhanced beyond these estimates by spectral binning. Thus, we find the features seen in our disk-integrated model could be detected in a reasonable amount of observing time, and JWST is poised to significantly enhance our view of disk ices and test the models presented here.

\section{Summary}
\label{sec:summary}

Motivated by advances in observational capabilities offered by JWST, SPHEREx, and future far-IR facilities, we present simulated ice observations of protoplanetary disks derived from realistic astrochemical models. We derive the abundances of ice species as a function of disk location and time and use these to create opacity mixtures that are incorporated into RADMC-3D to generate simulated disk images and spectra. Our main findings include: 

\begin{itemize}
    \item The generation of ice features involves extended background sources from both scattered light and direct thermal emission. Radiative transfer modeling is required to properly account for these effects when simulating observations. 
    \item Observations are most sensitive to ices above the disk midplane. Detailed chemical models, such as those we present here, are needed to connect the observable ices to a comprehensive understanding of ices throughout the entire disk.
    \item Near-to-mid-IR absorption features of H$_2$O, CO$_2$, and CH$_3$OH ice are readily detectable in disk-integrated spectra for inclinations $\gtrsim75^\circ$. The exact value will depend on the geometry of the disk, in particular the vertical structure and degree of flaring. Far-IR H$_2$O emission features are most prominent from disks closer to face-on.
    \item CO, CH$_4$, and NH$_3$ ice are not detectable in the disk-integrated spectra due to their concentration near the midplane (CO and CH$_4$), low abundance (CH$_4$ and NH$_3$), and intrinsically weak features (NH$_3$).
    \item Spatially resolved spectra of edge-on disks, as may be possible with JWST, offer more direct means to observe the vertical distribution of ice species.
    \item Spatially resolved spectra of face-on disks can trace ice features due to scattering, and may detect features from CO and CH$_4$ ice in the outer parts of the disk.
    \item CO$_2$ ice features show the strongest variation with time, increasing in strength from 0.25 to 2 Myr in the Reset models while decreasing in strength in the Inheritance models.
    \item CH$_3$OH has very low abundance in the Reset chemical models, resulting in no detectable features. Thus, observations targeting CH$_3$OH ice can discriminate between Inheritance and Reset scenarios.
\end{itemize}

We thank the referee for providing valuable feedback. N.P.B. and D.E.A.  acknowledge support from the UVA Virginia Initiative on Cosmic Origins. N.P.B. also acknowledges support from SOFIA grant 08-0101.  L.I.C. acknowledges support from the David and Lucille Packard Foundation, Johnson \& Johnson WISTEM2D, NSF AAG AST-1910106, and NASA ATP 80NSSC20K0529. We acknowledge the use of the SSHADE database \citep{SSHADE}, from which we accessed the NH$_3$ ice optical constants. We also gratefully acknowledge the use of UVA's Hydra Computing Cluster, which was utilized to run the radiative transfer models presented in this work.

\software{RADMC-3D \citep{dullemond2012}}

\bibliographystyle{aasjournal}
\bibliography{DiskIce}

\begin{thebibliography}{}
\expandafter\ifx\csname natexlab\endcsname\relax\def\natexlab#1{#1}\fi
\providecommand{\url}[1]{\href{#1}{#1}}
\providecommand{\dodoi}[1]{doi:~\href{http://doi.org/#1}{\nolinkurl{#1}}}
\providecommand{\doeprint}[1]{\href{http://ascl.net/#1}{\nolinkurl{http://ascl.net/#1}}}
\providecommand{\doarXiv}[1]{\href{https://arxiv.org/abs/#1}{\nolinkurl{https://arxiv.org/abs/#1}}}

\bibitem[{{Aikawa} {et~al.}(2012){Aikawa}, {Kamuro}, {Sakon}, {Itoh}, {Terada},
  {Noble}, {Pontoppidan}, {Fraser}, {Tamura}, {Kandori}, {Kawamura}, \&
  {Ueno}}]{aikawa2012}
{Aikawa}, Y., {Kamuro}, D., {Sakon}, I., {et~al.} 2012, \aap, 538, A57,
  \dodoi{10.1051/0004-6361/201015999}

\bibitem[{{Altwegg} {et~al.}(2019){Altwegg}, {Balsiger}, \&
  {Fuselier}}]{altwegg2019}
{Altwegg}, K., {Balsiger}, H., \& {Fuselier}, S.~A. 2019, \araa, 57, 113,
  \dodoi{10.1146/annurev-astro-091918-104409}

\bibitem[{{Anderson} {et~al.}(2021){Anderson}, {Blake}, {Cleeves}, {Bergin},
  {Zhang}, {Schwarz}, {Salyk}, \& {Bosman}}]{anderson2021}
{Anderson}, D.~E., {Blake}, G.~A., {Cleeves}, L.~I., {et~al.} 2021, \apj, 909,
  55, \dodoi{10.3847/1538-4357/abd9c1}

\bibitem[{{Andrews}(2020)}]{andrews2020}
{Andrews}, S.~M. 2020, \araa, 58, 483,
  \dodoi{10.1146/annurev-astro-031220-010302}

\bibitem[{{Andrews} {et~al.}(2011){Andrews}, {Wilner}, {Espaillat}, {Hughes},
  {Dullemond}, {McClure}, {Qi}, \& {Brown}}]{andrews2011}
{Andrews}, S.~M., {Wilner}, D.~J., {Espaillat}, C., {et~al.} 2011, \apj, 732,
  42, \dodoi{10.1088/0004-637X/732/1/42}

\bibitem[{{Baratta} \& {Palumbo}(1998)}]{baratta1998}
{Baratta}, G.~A., \& {Palumbo}, M.~E. 1998, Journal of the Optical Society of
  America A, 15, 3076, \dodoi{10.1364/JOSAA.15.003076}

\bibitem[{{Bergin} {et~al.}(2005){Bergin}, {Melnick}, {Gerakines}, {Neufeld},
  \& {Whittet}}]{bergin2005}
{Bergin}, E.~A., {Melnick}, G.~J., {Gerakines}, P.~A., {Neufeld}, D.~A., \&
  {Whittet}, D. C.~B. 2005, \apjl, 627, L33, \dodoi{10.1086/431932}

\bibitem[{{Bethell} \& {Bergin}(2011{\natexlab{a}})}]{bethell2011_photoe}
{Bethell}, T.~J., \& {Bergin}, E.~A. 2011{\natexlab{a}}, \apj, 740, 7,
  \dodoi{10.1088/0004-637X/740/1/7}

\bibitem[{{Bethell} \& {Bergin}(2011{\natexlab{b}})}]{bethell2011_lya}
---. 2011{\natexlab{b}}, \apj, 739, 78, \dodoi{10.1088/0004-637X/739/2/78}

\bibitem[{{Betti} {et~al.}(2021){Betti}, {Follette}, {Jorquera}, {Bonnefoy},
  {Perez}, {Chauvin}, {Boccaletti}, \& {Buenzli}}]{betti2021}
{Betti}, S., {Follette}, K., {Jorquera}, S., {et~al.} 2021, in American
  Astronomical Society Meeting Abstracts, Vol.~53, American Astronomical
  Society Meeting Abstracts, 325.02

\bibitem[{{Bohren} \& {Huffman}(1983)}]{bohren1983}
{Bohren}, C.~F., \& {Huffman}, D.~R. 1983, {Absorption and scattering of light
  by small particles} (New York: Wiley)

\bibitem[{{Boogert} {et~al.}(2015){Boogert}, {Gerakines}, \&
  {Whittet}}]{boogert2015}
{Boogert}, A.~C.~A., {Gerakines}, P.~A., \& {Whittet}, D. C.~B. 2015, \araa,
  53, 541, \dodoi{10.1146/annurev-astro-082214-122348}

\bibitem[{{Boogert} {et~al.}(2008){Boogert}, {Pontoppidan}, {Knez}, {Lahuis},
  {Kessler-Silacci}, {van Dishoeck}, {Blake}, {Augereau}, {Bisschop},
  {Bottinelli}, {Brooke}, {Brown}, {Crapsi}, {Evans}, {Fraser}, {Geers},
  {Huard}, {J{\o}rgensen}, {{\"O}berg}, {Allen}, {Harvey}, {Koerner}, {Mundy},
  {Padgett}, {Sargent}, \& {Stapelfeldt}}]{boogert2008}
{Boogert}, A.~C.~A., {Pontoppidan}, K.~M., {Knez}, C., {et~al.} 2008, \apj,
  678, 985, \dodoi{10.1086/533425}

\bibitem[{{Boogert} {et~al.}(2011){Boogert}, {Huard}, {Cook}, {Chiar}, {Knez},
  {Decin}, {Blake}, {Tielens}, \& {van Dishoeck}}]{boogert2011}
{Boogert}, A.~C.~A., {Huard}, T.~L., {Cook}, A.~M., {et~al.} 2011, \apj, 729,
  92, \dodoi{10.1088/0004-637X/729/2/92}

\bibitem[{{Bosman} {et~al.}(2018){Bosman}, {Walsh}, \& {van
  Dishoeck}}]{bosman2018}
{Bosman}, A.~D., {Walsh}, C., \& {van Dishoeck}, E.~F. 2018, \aap, 618, A182,
  \dodoi{10.1051/0004-6361/201833497}

\bibitem[{{Bouilloud} {et~al.}(2015){Bouilloud}, {Fray}, {B{\'e}nilan},
  {Cottin}, {Gazeau}, \& {Jolly}}]{bouilloud2015}
{Bouilloud}, M., {Fray}, N., {B{\'e}nilan}, Y., {et~al.} 2015, \mnras, 451,
  2145, \dodoi{10.1093/mnras/stv1021}

\bibitem[{{Calvet} \& {Gullbring}(1998)}]{calvet98}
{Calvet}, N., \& {Gullbring}, E. 1998, \apj, 509, 802, \dodoi{10.1086/306527}

\bibitem[{{Castelli} \& {Kurucz}(2003)}]{castelli2003}
{Castelli}, F., \& {Kurucz}, R.~L. 2003, in Modelling of Stellar Atmospheres,
  ed. N.~{Piskunov}, W.~W. {Weiss}, \& D.~F. {Gray}, Vol. 210, A20.
\newblock \doarXiv{astro-ph/0405087}

\bibitem[{{Cleeves} {et~al.}(2013){Cleeves}, {Adams}, \&
  {Bergin}}]{cleeves2013}
{Cleeves}, L.~I., {Adams}, F.~C., \& {Bergin}, E.~A. 2013, \apj, 772, 5,
  \dodoi{10.1088/0004-637X/772/1/5}

\bibitem[{{Cleeves} {et~al.}(2014){Cleeves}, {Bergin}, \&
  {Adams}}]{cleeves2014_cr}
{Cleeves}, L.~I., {Bergin}, E.~A., \& {Adams}, F.~C. 2014, \apj, 794, 123,
  \dodoi{10.1088/0004-637X/794/2/123}

\bibitem[{{Cleeves} {et~al.}(2015){Cleeves}, {Bergin}, {Qi}, {Adams}, \&
  {{\"O}berg}}]{cleeves2015_TWHya}
{Cleeves}, L.~I., {Bergin}, E.~A., {Qi}, C., {Adams}, F.~C., \& {{\"O}berg},
  K.~I. 2015, \apj, 799, 204, \dodoi{10.1088/0004-637X/799/2/204}

\bibitem[{{Cleeves} {et~al.}(2018){Cleeves}, {{\"O}berg}, {Wilner}, {Huang},
  {Loomis}, {Andrews}, \& {Guzman}}]{cleeves2018}
{Cleeves}, L.~I., {{\"O}berg}, K.~I., {Wilner}, D.~J., {et~al.} 2018, \apj,
  865, 155, \dodoi{10.3847/1538-4357/aade96}

\bibitem[{{Creech-Eakman} {et~al.}(2002){Creech-Eakman}, {Chiang}, {Joung},
  {Blake}, \& {van Dishoeck}}]{creech-eakman2002}
{Creech-Eakman}, M.~J., {Chiang}, E.~I., {Joung}, R.~M.~K., {Blake}, G.~A., \&
  {van Dishoeck}, E.~F. 2002, \aap, 385, 546,
  \dodoi{10.1051/0004-6361:20020157}

\bibitem[{{Dor{\'e}} {et~al.}(2018){Dor{\'e}}, {Werner}, {Ashby}, {Bleem},
  {Bock}, {Burt}, {Capak}, {Chang}, {Chaves-Montero}, {Chen}, {Civano},
  {Cleeves}, {Cooray}, {Crill}, {Crossfield}, {Cushing}, {de la Torre},
  {DiMatteo}, {Dvory}, {Dvorkin}, {Espaillat}, {Ferraro}, {Finkbeiner},
  {Greene}, {Hewitt}, {Hogg}, {Huffenberger}, {Jun}, {Ilbert}, {Jeong},
  {Johnson}, {Kim}, {Kirkpatrick}, {Kowalski}, {Korngut}, {Li}, {Lisse},
  {MacGregor}, {Mamajek}, {Mauskopf}, {Melnick}, {M{\'e}nard}, {Neyrinck},
  {{\"O}berg}, {Pisani}, {Rocca}, {Salvato}, {Schaan}, {Scoville}, {Song},
  {Stevens}, {Tenneti}, {Teplitz}, {Tolls}, {Unwin}, {Urry}, {Wandelt},
  {Williams}, {Wilner}, {Windhorst}, {Wolk}, {Yorke}, \& {Zemcov}}]{dore2018}
{Dor{\'e}}, O., {Werner}, M.~W., {Ashby}, M. L.~N., {et~al.} 2018, arXiv
  e-prints, arXiv:1805.05489.
\newblock \doarXiv{1805.05489}

\bibitem[{{Draine}(2003)}]{draine2003}
{Draine}, B.~T. 2003, \apj, 598, 1026, \dodoi{10.1086/379123}

\bibitem[{{Dra{\.z}kowska} \& {Alibert}(2017)}]{drazkowska2017}
{Dra{\.z}kowska}, J., \& {Alibert}, Y. 2017, \aap, 608, A92,
  \dodoi{10.1051/0004-6361/201731491}

\bibitem[{{Dullemond} {et~al.}(2012){Dullemond}, {Juhasz}, {Pohl}, {Sereshti},
  {Shetty}, {Peters}, {Commercon}, \& {Flock}}]{dullemond2012}
{Dullemond}, C.~P., {Juhasz}, A., {Pohl}, A., {et~al.} 2012, {RADMC-3D: A
  multi-purpose radiative transfer tool}, Astrophysics Source Code Library.
\newblock \doeprint{1202.015}

\bibitem[{{Evans} {et~al.}(2009){Evans}, {Dunham}, {J{\o}rgensen}, {Enoch},
  {Mer{\'\i}n}, {van Dishoeck}, {Alcal{\'a}}, {Myers}, {Stapelfeldt}, {Huard},
  {Allen}, {Harvey}, {van Kempen}, {Blake}, {Koerner}, {Mundy}, {Padgett}, \&
  {Sargent}}]{evans2009}
{Evans}, Neal~J., I., {Dunham}, M.~M., {J{\o}rgensen}, J.~K., {et~al.} 2009,
  \apjs, 181, 321, \dodoi{10.1088/0067-0049/181/2/321}

\bibitem[{{Facchini} {et~al.}(2021){Facchini}, {Teague}, {Bae}, {Benisty},
  {Keppler}, \& {Isella}}]{facchini2021}
{Facchini}, S., {Teague}, R., {Bae}, J., {et~al.} 2021, \aj, 162, 99,
  \dodoi{10.3847/1538-3881/abf0a4}

\bibitem[{{Feigelson} \& {Montmerle}(1999)}]{feigelson1999}
{Feigelson}, E.~D., \& {Montmerle}, T. 1999, \araa, 37, 363,
  \dodoi{10.1146/annurev.astro.37.1.363}

\bibitem[{{Fogel} {et~al.}(2011){Fogel}, {Bethell}, {Bergin}, {Calvet}, \&
  {Semenov}}]{fogel2011}
{Fogel}, J. K.~J., {Bethell}, T.~J., {Bergin}, E.~A., {Calvet}, N., \&
  {Semenov}, D. 2011, \apj, 726, 29, \dodoi{10.1088/0004-637X/726/1/29}

\bibitem[{{Giuliano} {et~al.}(2019){Giuliano}, {Gavdush}, {M{\"u}ller},
  {Zaytsev}, {Grassi}, {Ivlev}, {Palumbo}, {Baratta}, {Scir{\`e}}, {Komand in},
  {Yurchenko}, \& {Caselli}}]{giuliano2019}
{Giuliano}, B.~M., {Gavdush}, A.~A., {M{\"u}ller}, B., {et~al.} 2019, \aap,
  629, A112, \dodoi{10.1051/0004-6361/201935619}

\bibitem[{{Gomes} {et~al.}(2005){Gomes}, {Levison}, {Tsiganis}, \&
  {Morbidelli}}]{gomes2005}
{Gomes}, R., {Levison}, H.~F., {Tsiganis}, K., \& {Morbidelli}, A. 2005,
  Nature, 435, 466, \dodoi{10.1038/nature03676}

\bibitem[{{Gullbring} {et~al.}(2000){Gullbring}, {Calvet}, {Muzerolle}, \&
  {Hartmann}}]{gullbring2000}
{Gullbring}, E., {Calvet}, N., {Muzerolle}, J., \& {Hartmann}, L. 2000, \apj,
  544, 927, \dodoi{10.1086/317253}

\bibitem[{{Gundlach} \& {Blum}(2015)}]{gundlach2015}
{Gundlach}, B., \& {Blum}, J. 2015, \apj, 798, 34,
  \dodoi{10.1088/0004-637X/798/1/34}

\bibitem[{{Harada} {et~al.}(2010){Harada}, {Herbst}, \& {Wakelam}}]{harada2010}
{Harada}, N., {Herbst}, E., \& {Wakelam}, V. 2010, \apj, 721, 1570,
  \dodoi{10.1088/0004-637X/721/2/1570}

\bibitem[{{Harries}(2000)}]{harries2000}
{Harries}, T.~J. 2000, \mnras, 315, 722,
  \dodoi{10.1046/j.1365-8711.2000.03505.x}

\bibitem[{{Harries} {et~al.}(2004){Harries}, {Monnier}, {Symington}, \&
  {Kurosawa}}]{harries2004}
{Harries}, T.~J., {Monnier}, J.~D., {Symington}, N.~H., \& {Kurosawa}, R. 2004,
  \mnras, 350, 565, \dodoi{10.1111/j.1365-2966.2004.07668.x}

\bibitem[{{Hartmann} {et~al.}(1998){Hartmann}, {Calvet}, {Gullbring}, \&
  {D'Alessio}}]{hartmann1998}
{Hartmann}, L., {Calvet}, N., {Gullbring}, E., \& {D'Alessio}, P. 1998, \apj,
  495, 385, \dodoi{10.1086/305277}

\bibitem[{{Haworth} {et~al.}(2016){Haworth}, {Ilee}, {Forgan}, {Facchini},
  {Price}, {Boneberg}, {Booth}, {Clarke}, {Gonzalez}, {Hutchison}, {Kamp},
  {Laibe}, {Lyra}, {Meru}, {Mohanty}, {Pani{\'c}}, {Rice}, {Suzuki}, {Teague},
  {Walsh}, {Woitke}, \& {Community authors}}]{haworth2016}
{Haworth}, T.~J., {Ilee}, J.~D., {Forgan}, D.~H., {et~al.} 2016, \pasa, 33,
  e053, \dodoi{10.1017/pasa.2016.45}

\bibitem[{{Heays} {et~al.}(2017){Heays}, {Bosman}, \& {van
  Dishoeck}}]{heays2017}
{Heays}, A.~N., {Bosman}, A.~D., \& {van Dishoeck}, E.~F. 2017, \aap, 602,
  A105, \dodoi{10.1051/0004-6361/201628742}

\bibitem[{{Honda} {et~al.}(2009){Honda}, {Inoue}, {Fukagawa}, {Oka},
  {Nakamoto}, {Ishii}, {Terada}, {Takato}, {Kawakita}, {Okamoto}, {Shibai},
  {Tamura}, {Kudo}, \& {Itoh}}]{honda2009}
{Honda}, M., {Inoue}, A.~K., {Fukagawa}, M., {et~al.} 2009, ApJ, 690, L110,
  \dodoi{10.1088/0004-637X/690/2/L110}

\bibitem[{{Honda} {et~al.}(2016){Honda}, {Kudo}, {Takatsuki}, {Inoue},
  {Nakamoto}, {Fukagawa}, {Tamura}, {Terada}, \& {Takato}}]{honda2016}
{Honda}, M., {Kudo}, T., {Takatsuki}, S., {et~al.} 2016, \apj, 821, 2,
  \dodoi{10.3847/0004-637X/821/1/2}

\bibitem[{{Hudgins} {et~al.}(1993){Hudgins}, {Sandford}, {Allamandola}, \&
  {Tielens}}]{hudgins1993}
{Hudgins}, D.~M., {Sandford}, S.~A., {Allamandola}, L.~J., \& {Tielens},
  A.~G.~G.~M. 1993, \apjs, 86, 713, \dodoi{10.1086/191796}

\bibitem[{{Hudson}(2020)}]{HudsonWebsite}
{Hudson}, R.~L. 2020, The Cosmic Ice Laboratory,
  \url{https://science.gsfc.nasa.gov/691/cosmicice/}

\bibitem[{{Inoue} {et~al.}(2008){Inoue}, {Honda}, {Nakamoto}, \&
  {Oka}}]{inoue2008}
{Inoue}, A.~K., {Honda}, M., {Nakamoto}, T., \& {Oka}, A. 2008, \pasj, 60, 557,
  \dodoi{10.1093/pasj/60.3.557}

\bibitem[{{Jewitt} \& {Luu}(2004)}]{jewitt2004}
{Jewitt}, D.~C., \& {Luu}, J. 2004, \nat, 432, 731, \dodoi{10.1038/nature03111}

\bibitem[{{Kamp} {et~al.}(2018){Kamp}, {Scheepstra}, {Min}, {Klarmann}, \&
  {Riviere-Marichalar}}]{kamp2018}
{Kamp}, I., {Scheepstra}, A., {Min}, M., {Klarmann}, L., \&
  {Riviere-Marichalar}, P. 2018, \aap, 617, A1,
  \dodoi{10.1051/0004-6361/201732368}

\bibitem[{{Krijt} {et~al.}(2020){Krijt}, {Bosman}, {Zhang}, {Schwarz},
  {Ciesla}, \& {Bergin}}]{krijt2020}
{Krijt}, S., {Bosman}, A.~D., {Zhang}, K., {et~al.} 2020, \apj, 899, 134,
  \dodoi{10.3847/1538-4357/aba75d}

\bibitem[{{Kurosawa} {et~al.}(2004){Kurosawa}, {Harries}, {Bate}, \&
  {Symington}}]{kurosawa2004}
{Kurosawa}, R., {Harries}, T.~J., {Bate}, M.~R., \& {Symington}, N.~H. 2004,
  \mnras, 351, 1134, \dodoi{10.1111/j.1365-2966.2004.07869.x}

\bibitem[{{Lynden-Bell} \& {Pringle}(1974)}]{lynden-bell1974}
{Lynden-Bell}, D., \& {Pringle}, J.~E. 1974, \mnras, 168, 603,
  \dodoi{10.1093/mnras/168.3.603}

\bibitem[{{Malfait} {et~al.}(1999){Malfait}, {Waelkens}, {Bouwman}, {de Koter},
  \& {Waters}}]{malfait1999}
{Malfait}, K., {Waelkens}, C., {Bouwman}, J., {de Koter}, A., \& {Waters},
  L.~B.~F.~M. 1999, \aap, 345, 181

\bibitem[{{Malfait} {et~al.}(1998){Malfait}, {Waelkens}, {Waters},
  {Vandenbussche}, {Huygen}, \& {de Graauw}}]{malfait1998}
{Malfait}, K., {Waelkens}, C., {Waters}, L.~B.~F.~M., {et~al.} 1998, \aap, 332,
  L25

\bibitem[{{Mathis} {et~al.}(1977){Mathis}, {Rumpl}, \&
  {Nordsieck}}]{mathis1977}
{Mathis}, J.~S., {Rumpl}, W., \& {Nordsieck}, K.~H. 1977, \apj, 217, 425,
  \dodoi{10.1086/155591}

\bibitem[{{Matr{\`a}} {et~al.}(2018){Matr{\`a}}, {Wilner}, {{\"O}berg},
  {Andrews}, {Loomis}, {Wyatt}, \& {Dent}}]{matra2018}
{Matr{\`a}}, L., {Wilner}, D.~J., {{\"O}berg}, K.~I., {et~al.} 2018, \apj, 853,
  147, \dodoi{10.3847/1538-4357/aaa42a}

\bibitem[{{Matr{\`a}} {et~al.}(2017){Matr{\`a}}, {MacGregor}, {Kalas}, {Wyatt},
  {Kennedy}, {Wilner}, {Duchene}, {Hughes}, {Pan}, {Shannon}, {Clampin},
  {Fitzgerald}, {Graham}, {Holland}, {Pani{\'c}}, \& {Su}}]{matra2017}
{Matr{\`a}}, L., {MacGregor}, M.~A., {Kalas}, P., {et~al.} 2017, \apj, 842, 9,
  \dodoi{10.3847/1538-4357/aa71b4}

\bibitem[{{McCabe} {et~al.}(2011){McCabe}, {Duch{\^e}ne}, {Pinte},
  {Stapelfeldt}, {Ghez}, \& {M{\'e}nard}}]{mccabe2011}
{McCabe}, C., {Duch{\^e}ne}, G., {Pinte}, C., {et~al.} 2011, \apj, 727, 90,
  \dodoi{10.1088/0004-637X/727/2/90}

\bibitem[{{McClure} {et~al.}(2015){McClure}, {Espaillat}, {Calvet}, {Bergin},
  {D'Alessio}, {Watson}, {Manoj}, {Sargent}, \& {Cleeves}}]{mcclure2015}
{McClure}, M.~K., {Espaillat}, C., {Calvet}, N., {et~al.} 2015, \apj, 799, 162,
  \dodoi{10.1088/0004-637X/799/2/162}

\bibitem[{{Meyer} {et~al.}(1997){Meyer}, {Cardelli}, \& {Sofia}}]{meyer1997}
{Meyer}, D.~M., {Cardelli}, J.~A., \& {Sofia}, U.~J. 1997, \apjl, 490, L103,
  \dodoi{10.1086/311023}

\bibitem[{{Min} {et~al.}(2016){Min}, {Bouwman}, {Dominik}, {Waters},
  {Pontoppidan}, {Hony}, {Mulders}, {Henning}, {van Dishoeck}, {Woitke},
  {Evans}, \& {Digit Team}}]{min2016}
{Min}, M., {Bouwman}, J., {Dominik}, C., {et~al.} 2016, \aap, 593, A11,
  \dodoi{10.1051/0004-6361/201425432}

\bibitem[{{Mumma} \& {Charnley}(2011)}]{mumma2011}
{Mumma}, M.~J., \& {Charnley}, S.~B. 2011, \araa, 49, 471,
  \dodoi{10.1146/annurev-astro-081309-130811}

\bibitem[{{Musiolik} {et~al.}(2016){Musiolik}, {Teiser}, {Jankowski}, \&
  {Wurm}}]{musiolik2016}
{Musiolik}, G., {Teiser}, J., {Jankowski}, T., \& {Wurm}, G. 2016, \apj, 818,
  16, \dodoi{10.3847/0004-637X/818/1/16}

\bibitem[{{{\"O}berg} \& {Bergin}(2021)}]{oberg2021_review}
{{\"O}berg}, K.~I., \& {Bergin}, E.~A. 2021, \physrep, 893, 1,
  \dodoi{10.1016/j.physrep.2020.09.004}

\bibitem[{{{\"O}berg} {et~al.}(2011){{\"O}berg}, {Boogert}, {Pontoppidan}, {van
  den Broek}, {van Dishoeck}, {Bottinelli}, {Blake}, \& {Evans}}]{oberg2011}
{{\"O}berg}, K.~I., {Boogert}, A.~C.~A., {Pontoppidan}, K.~M., {et~al.} 2011,
  \apj, 740, 109, \dodoi{10.1088/0004-637X/740/2/109}

\bibitem[{{{\"O}berg} {et~al.}(2021){{\"O}berg}, {Cleeves}, {Bergner},
  {Cavanaro}, {Teague}, {Huang}, {Loomis}, {Bergin}, {Blake}, {Calahan},
  {Cazzoletti}, {Guzm{\'a}n}, {Hogerheijde}, {Kama}, {Terwisscha van
  Scheltinga}, {Qi}, {van Dishoeck}, {Walsh}, \& {Wilner}}]{oberg2021}
{{\"O}berg}, K.~I., {Cleeves}, L.~I., {Bergner}, J.~B., {et~al.} 2021, \aj,
  161, 38, \dodoi{10.3847/1538-3881/abc74d}

\bibitem[{{Oka} {et~al.}(2012){Oka}, {Inoue}, {Nakamoto}, \& {Honda}}]{oka2012}
{Oka}, A., {Inoue}, A.~K., {Nakamoto}, T., \& {Honda}, M. 2012, \apj, 747, 138,
  \dodoi{10.1088/0004-637X/747/2/138}

\bibitem[{{Palumbo} {et~al.}(2006){Palumbo}, {Baratta}, {Collings}, \&
  {McCoustra}}]{palumbo2006}
{Palumbo}, M.~E., {Baratta}, G.~A., {Collings}, M.~P., \& {McCoustra}, M.~R.~S.
  2006, Physical Chemistry Chemical Physics (Incorporating Faraday
  Transactions), 8, 279

\bibitem[{{Pinte} {et~al.}(2009){Pinte}, {Harries}, {Min}, {Watson},
  {Dullemond}, {Woitke}, {M{\'e}nard}, \& {Dur{\'a}n-Rojas}}]{pinte2009}
{Pinte}, C., {Harries}, T.~J., {Min}, M., {et~al.} 2009, \aap, 498, 967,
  \dodoi{10.1051/0004-6361/200811555}

\bibitem[{{Pontoppidan} {et~al.}(2005){Pontoppidan}, {Dullemond}, {van
  Dishoeck}, {Blake}, {Boogert}, {Evans}, {Kessler-Silacci}, \&
  {Lahuis}}]{pontoppidan2005}
{Pontoppidan}, K.~M., {Dullemond}, C.~P., {van Dishoeck}, E.~F., {et~al.} 2005,
  \apj, 622, 463, \dodoi{10.1086/427688}

\bibitem[{{Pontoppidan} {et~al.}(2014){Pontoppidan}, {Salyk}, {Bergin},
  {Brittain}, {Marty}, {Mousis}, \& {{\"O}berg}}]{pontoppidan2014}
{Pontoppidan}, K.~M., {Salyk}, C., {Bergin}, E.~A., {et~al.} 2014, in
  Protostars and Planets VI, ed. H.~{Beuther}, R.~S. {Klessen}, C.~P.
  {Dullemond}, \& T.~{Henning}, 363,
  \dodoi{10.2458/azu_uapress_9780816531240-ch016}

\bibitem[{{Preibisch} {et~al.}(2005){Preibisch}, {Kim}, {Favata}, {Feigelson},
  {Flaccomio}, {Getman}, {Micela}, {Sciortino}, {Stassun}, {Stelzer}, \&
  {Zinnecker}}]{preibisch2005}
{Preibisch}, T., {Kim}, Y.-C., {Favata}, F., {et~al.} 2005, \apjs, 160, 401,
  \dodoi{10.1086/432891}

\bibitem[{{Price} {et~al.}(2020){Price}, {Cleeves}, \& {{\"O}berg}}]{price2020}
{Price}, E.~M., {Cleeves}, L.~I., \& {{\"O}berg}, K.~I. 2020, \apj, 890, 154,
  \dodoi{10.3847/1538-4357/ab5fd4}

\bibitem[{{Raymond} \& {Izidoro}(2017)}]{raymond2017}
{Raymond}, S.~N., \& {Izidoro}, A. 2017, Icarus, 297, 134,
  \dodoi{10.1016/j.icarus.2017.06.030}

\bibitem[{{Ripple} {et~al.}(2013){Ripple}, {Heyer}, {Gutermuth}, {Snell}, \&
  {Brunt}}]{ripple2013}
{Ripple}, F., {Heyer}, M.~H., {Gutermuth}, R., {Snell}, R.~L., \& {Brunt},
  C.~M. 2013, \mnras, 431, 1296, \dodoi{10.1093/mnras/stt247}

\bibitem[{{Schmitt} {et~al.}(2018){Schmitt}, {Bollard}, {Albert}, {Garenne},
  {Gorbacheva}, {Bonal}, {Volcke}, \& {the SSHADE partner's
  consortium}}]{SSHADE}
{Schmitt}, B., {Bollard}, P., {Albert}, D., {et~al.} 2018, {SSHADE: ``Solid
  Spectroscopy Hosting Architecture of Databases and Expertise" and its
  databases.},  OSUG Data Center. Service/Database Infrastructure.,
  \dodoi{10.26302/SSHADE}

\bibitem[{{Schwarz} {et~al.}(2018){Schwarz}, {Bergin}, {Cleeves}, {Zhang},
  {{\"O}berg}, {Blake}, \& {Anderson}}]{schwarz2018}
{Schwarz}, K.~R., {Bergin}, E.~A., {Cleeves}, L.~I., {et~al.} 2018, \apj, 856,
  85, \dodoi{10.3847/1538-4357/aaae08}

\bibitem[{{Seifert} {et~al.}(2021){Seifert}, {Cleeves}, {Adams}, \&
  {Li}}]{seifert2021}
{Seifert}, R.~A., {Cleeves}, L.~I., {Adams}, F.~C., \& {Li}, Z.-Y. 2021, \apj,
  912, 136, \dodoi{10.3847/1538-4357/abf09a}

\bibitem[{{Stevenson} \& {Lunine}(1988)}]{stevenson1988}
{Stevenson}, D.~J., \& {Lunine}, J.~I. 1988, \icarus, 75, 146,
  \dodoi{10.1016/0019-1035(88)90133-9}

\bibitem[{{Str{\o}m} {et~al.}(2020){Str{\o}m}, {Bodewits}, {Knight}, {Kiefer},
  {Jones}, {Kral}, {Matr{\`a}}, {Bodman}, {Capria}, {Cleeves}, {Fitzsimmons},
  {Haghighipour}, {Harrison}, {Iglesias}, {Kama}, {Linnartz}, {Majumdar}, {de
  Mooij}, {Milam}, {Opitom}, {Rebollido}, {Rogers}, {Snodgrass}, {Sousa-Silva},
  {Xu}, {Lin}, \& {Zieba}}]{strom2020}
{Str{\o}m}, P.~A., {Bodewits}, D., {Knight}, M.~M., {et~al.} 2020, \pasp, 132,
  101001, \dodoi{10.1088/1538-3873/aba6a0}

\bibitem[{{Terada} \& {Tokunaga}(2012)}]{terada2012}
{Terada}, H., \& {Tokunaga}, A.~T. 2012, \apj, 753, 19,
  \dodoi{10.1088/0004-637X/753/1/19}

\bibitem[{{Terada} \& {Tokunaga}(2017)}]{terada2017}
---. 2017, \apj, 834, 115, \dodoi{10.3847/1538-4357/834/2/115}

\bibitem[{{Terada} {et~al.}(2007){Terada}, {Tokunaga}, {Kobayashi}, {Takato},
  {Hayano}, \& {Takami}}]{terada2007}
{Terada}, H., {Tokunaga}, A.~T., {Kobayashi}, N., {et~al.} 2007, \apj, 667,
  303, \dodoi{10.1086/520951}

\bibitem[{{Trotta}(1996)}]{trotta1996}
{Trotta}, F. 1996, PhD thesis, Joseph Fourier University

\bibitem[{{van den Ancker} {et~al.}(2000){van den Ancker}, {Bouwman},
  {Wesselius}, {Waters}, {Dougherty}, \& {van Dishoeck}}]{vandenancker2000}
{van den Ancker}, M.~E., {Bouwman}, J., {Wesselius}, P.~R., {et~al.} 2000,
  \aap, 357, 325.
\newblock \doarXiv{astro-ph/0002440}

\bibitem[{{Wang} {et~al.}(2005){Wang}, {Bell}, {Iedema}, {Tsekouras}, \&
  {Cowin}}]{wang2005}
{Wang}, H., {Bell}, R.~C., {Iedema}, M.~J., {Tsekouras}, A.~A., \& {Cowin},
  J.~P. 2005, \apj, 620, 1027, \dodoi{10.1086/427072}

\bibitem[{{Warren} \& {Brandt}(2008)}]{warren2008}
{Warren}, S.~G., \& {Brandt}, R.~E. 2008, JGRD, 113, 14220,
  \dodoi{doi:10.1029/2007JD009744}

\bibitem[{{Whittet} {et~al.}(1983){Whittet}, {Bode}, {Longmore}, {Baines}, \&
  {Evans}}]{whittet1983}
{Whittet}, D.~C.~B., {Bode}, M.~F., {Longmore}, A.~J., {Baines}, D.~W.~T., \&
  {Evans}, A. 1983, \nat, 303, 218, \dodoi{10.1038/303218a0}

\bibitem[{{Woitke} {et~al.}(2019){Woitke}, {Kamp}, {Antonellini}, {Anthonioz},
  {Baldovin-Saveedra}, {Carmona}, {Dionatos}, {Dominik}, {Greaves},
  {G{\"u}del}, {Ilee}, {Liebhardt}, {Menard}, {Min}, {Pinte}, {Rab}, {Rigon},
  {Thi}, {Thureau}, \& {Waters}}]{woitke2019}
{Woitke}, P., {Kamp}, I., {Antonellini}, S., {et~al.} 2019, \pasp, 131, 064301,
  \dodoi{10.1088/1538-3873/aaf4e5}

\end{thebibliography}

\end{document}